\documentclass[aps,prb,10pt,twocolumn,amsmath,amssymb,longbibliography,superscriptaddress,nofootinbib]{revtex4-2}
\usepackage{amsmath,amssymb,amsfonts} 	
\usepackage{graphicx}
\usepackage{hhline}
\usepackage[latin1]{inputenc}
\usepackage{subfigure}
\usepackage[normalem]{ulem}
\usepackage{xcolor}
\usepackage{multirow}
\begin{document}

\title{
Classifying magnons in itinerant ferromagnets from linear response TDDFT: Fe, Ni and Co revisited.}

\author{Thorbj\o rn Skovhus}
\email{thosk@dtu.dk}
\affiliation{Department of Physics and Astronomy, Uppsala University, 
Box 516, 
751 20 Uppsala, Sweden}
\affiliation{CAMD, Department of Physics, Technical University of Denmark, 2800 Kgs. Lyngby, Denmark}
\author{Thomas Olsen}
\affiliation{CAMD, Department of Physics, Technical University of Denmark, 2800 Kgs. Lyngby, Denmark}

\begin{abstract}
The magnetic excitation spectrum of itinerant magnets exhibits rich and complex spectral features that often complicate interpretation of the underlying physics. For perturbations in the long wavelength limit, one obtains a well defined pole at zero frequency in the spectral function, the Goldstone magnon. However, for optical modes and finite wavevectors, the magnon spectrum may become damped, exhibit branching, or be completely washed out. In the present work, we show how the physical mechanism of all such features can be understood from careful analysis of the eigenmodes of the many-body spectral function. We perform first principles computations of elemental itinerant ferromagnets using a novel implementation of the linear response time-dependent density functional theory (LR-TDDFT) framework and classify the collective nature of individual spectral features based on the self-enhancement function, the product of the noninteracting Kohn-Sham susceptibility and the exchange-correlation kernel. 
In particular, we distinguish between coherent and incoherent collective excitations, depending on whether the real part of the self-enhancement function crosses unity at the spectral peak of the magnon,
which may or may not be subject to Landau damping as quantified by the imaginary part. 
Classifying the computed magnon spectra accordingly, we observe coexistence of coherent magnon branches in bcc-Fe, as well as decoherence of the primary magnon branch in fcc-Ni for wave vectors near the BZ boundary where incoherent valley magnons instead carry substantial spectral weight. 
The analysis also naturally leads to a 
definition of the many-body Stoner spectrum and allows us to quantify the binding energy of the Stoner pair excitations.
\end{abstract}
\maketitle

\section{Introduction}\label{sec:intro}
The low-energy excitations in magnetic materials are observable as poles in the dynamic magnetic susceptibility. For pristine insulators, the poles are undamped and interpreted as collective quasi-particle excitations, known as magnons \cite{boothroyd_principles_2020}. In such cases, the magnetic properties can often be modeled rather accurately by a Heisenberg Hamiltonian including a few nearest neighbor exchange interactions. The basic excitation spectrum may then be obtained within linear spin-wave theory, where the magnons emerge naturally as (approximate) bosonic excitations carrying a well-defined energy and momentum \cite{yosida_theory_1996}. Beyond linear spin-wave theory, magnons interact, resulting in broadening of the energy-momentum dispersion \cite{oguchi_theory_1960}, but such interactions are often negligible at low temperatures, and magnons may be regarded as approximate eigenstates of the full many-body Hamiltonian. 

For typical (non-half-metallic) itinerant magnets, the situation is much less clear. Such materials allow for Stoner (single-particle) excitations at arbitrarily low energies, and these may couple to the collective excitations and result in more complex spectral features \cite{Moriya1985, Friedrich2020}. In the simplest case, the Stoner continuum will give rise to a Lorentzian broadening of the magnon peak, which typically increases for increasing wavevectors \cite{Buczek2009}. The position of the spectral peak may then still be interpreted as a magnon energy, the broadening corresponding to a finite lifetime. However, the coupling itself may also give rise to distinct peaks or branching points in the spectral function \cite{Friedrich2020, Skovhus2024}, and in such cases the notions of collective excitations or magnons become unclear. It is not possible to incorporate the coupling to Stoner excitations in localized models (such as the Heisenberg Hamiltonian) in a simple manner, and it is natural to ask whether the concept of magnons is even well defined in itinerant magnets. On the other hand, in the absence of spin-orbit coupling, the magnetic order (in itinerant as well as insulating magnets) originates from spontaneous symmetry breaking and the excitation spectrum is guaranteed to contain a massless Goldstone boson \cite{goldstone_field_1961} that corresponds to a global rotation of the magnetic dipole density. Magnetic order is therefore always associated with the existence of at least one collective excitation in the long wavelength limit, which may be interpreted as a magnon. The question that remains is to what extent such an interpretation remains valid at finite wavevectors.

While the magnetic excitations in itinerant magnets are not expected to be captured well by site-based models, it is nevertheless possible to perform a well-defined adiabatic mapping of first principles total energy calculations to a Heisenberg model \cite{halilov_adiabatic_1998,Szilva2023}. Solving the model (in the linear spin-wave approximation) yields magnon eigenvalues at all wave vectors by construction, but such calculations are only expected to be reliable for the acoustic Goldstone magnon in the long wavelength limit \cite{Durhuus2023}. At generic wave vectors, the adiabatic magnon spectra at best represent weighted averages of the spectral function, and the interpretation of such results as proper magnon dispersions is dubious. Proper treatment of magnetic excitations in itinerant systems requires an analysis of the full spectral function (as a function of frequency) at a given wave vector. This can be done with first principles methods using the frameworks of either dynamical mean-field theory \cite{kotliar_electronic_2006}, many-body perturbation theory \cite{Aryasetiawan1999,Karlsson2000,SasIoglu2010, Muller2016,Okumura2019,olsen_unified_2021} or time-dependent density functional theory (TDDFT) \cite{Savrasov1998,Buczek2011b,Lounis2011,Rousseau2012,Cao2017,Singh2019,Tancogne-Dejean2020,Skovhus2021,Liu2023,binci_magnons_2025}. In order to obtain an accurate description of the coupling between collective modes and the Stoner continuum, it is crucial that the methodology reproduces the Goldstone condition: The energy of the lowest collective mode should approach zero in the long wave length limit. From the above considerations, this may appear to be a rather simple matter since it is a question of symmetry. However, in practical applications, it is often hard to satisfy the Goldstone condition. For example, many-body approaches typically yield a large Goldstone gap due to inconsistencies between single-particle and two-particle descriptions \cite{Muller2016}, and TDDFT approaches often produce a finite gap due to numerical inconsistencies in the implementation \cite{Buczek2011b,Lounis2011,Rousseau2012,Skovhus2021,Skovhus2022a,Skovhus2022b}. The fundamental magnon dispersion can often be well reproduced by performing a rigid shift of the magnon energies to accommodate the Goldstone criterion. However, this implies that the magnons are not positioned correctly with respect to the Stoner continuum and that the coupling may not be accurately captured.

In the present work, we reformulate the framework of linear response time-dependent density functional theory (LR-TDDFT) in such a way that all many-body correlations effects are contained within a single object, the self-enhancement function. By implementing computation of the self-enhancement function directly within the projector-augmented wave (PAW) method, we circumvent the numerical inconsistencies of previous approaches, enabling us to eliminate the Goldstone gap error through raw numerical convergence. We then move on and analyze the spectral functions of representative itinerant magnets (bcc-Fe, fcc-Ni, fcc-Co and hcp-Co) in detail. The present approach is ideally suited for this, since the self-enhancement function allows us to directly assess the collective character of specific features in the spectral function. Performing an eigenmode decomposition of the spectral function, we separate collective and single-particle spectral contributions and classify the collective character of individual magnon resonances. From the eigenmodes of the self-enhancement function, we also unravel the origin of complex spectral features such as Stoner stripes and valley magnons. 

The paper is organized as follows. In Sec. \ref{sec:theory} we introduce the basic theory underlying the LR-TDDFT approach, highlighting the appearance of the self-enhancement function, which governs the collective character of excitations. In Sec. \ref{sec:implementation} we document our new PAW implementation and demonstrate that it allows us to strictly eliminate the gap error in the limit of large basis sets. In Sec. \ref{sec:transition metals results} we present and discuss our results for bcc-Fe, fcc-Ni, fcc-Co, and hcp-Co. We start by introducing the eigenmode decomposition and then apply it to unravel the nature of magnetic excitations in these four prototypical compounds. In Sec. \ref{sec:conclusion} we provide conclusions and an outlook.


\section{Theory}\label{sec:theory}

\subsection{Transverse magnetic excitations in linear response theory}\label{sec:linear response theory}

For any quantum system, one can hierarchically order the excited states $|\alpha\rangle$ relative to a reference ground state $|0\rangle$ and single-particle operator $\hat{A}$ based on the lowest polynomial order $n$ which yields a finite excited-state overlap $\langle \alpha|\hat{A}^n|0\rangle \neq 0$. In this sense, the set of excited states with linear order ($n=1$) overlaps have the shortest hyperdistance to the ground state in the Hilbert space of the system. More importantly, these are also the excitations that dominate both inherent fluctuations at low temperatures as well as the response to weak external perturbations \cite{Nyquist1928,Callen1951,Kubo1957,Kubo1966}. 

For a system of electrons in particular, the necessary internal coordinates (single-particle operators) to consider are the four components of the electron density $n^\mu(\mathbf{r})$ where $\mu\in\{0,x,y,z\}$, as formalized in density functional theory \cite{HohenbergP.1973,Kohn1965} and its extension to time-dependent external potentials \cite{Runge1984}. 
Here, perturbations enter as external electromagnetic fields in the form of a scalar potential and a magnetic field, $W_\mathrm{ext}^\mu = \left(-e \phi_\mathrm{ext}, \mu_\mathrm{B} \mathbf{B}_\mathrm{ext} \right)$, to which the electronic system's linear order response is given by the four-component susceptibility tensor $\chi^{\mu\nu}$,
\begin{equation}
    \delta n^\mu(\mathbf{r}, t) = \sum_\nu \int_{-\infty}^\infty dt' \int d\mathbf{r}'\, \chi^{\mu\nu}(\mathbf{r}, \mathbf{r}', t-t') W_\mathrm{ext}^\nu(\mathbf{r}', t').
    \label{eq:linear response relation}
\end{equation}
%
In the zero temperature limit, the susceptibility $\chi^{\mu\nu}$ is composed of poles in frequency space corresponding to the excitation energies of states with linear order overlaps in $\hat{n}^\mu(\mathbf{r})$ and $\hat{n}^\nu(\mathbf{r}')$. By probing $\chi^{\mu\nu}$ theoretically or experimentally one can therefore directly characterize the order $n=1$ excitations of the electronic system.

The role of transverse magnetic excitations in such a characterization is most clearly illustrated for collinear magnetic systems absent of spin-orbit coupling. In this case, the total electronic spin projection along the magnetization axis ($S^z$) can be taken as a good quantum number and $\chi^{\mu\nu}$ becomes block diagonal, separating the transverse $\{x,y\}$ components from the longitudinal $\{0,z\}$ components, see e.g. \cite{Buczek2011b,Skovhus2021}. The transverse block is diagonalized by the retarded circular coordinate susceptibilities $\chi^{+-}$ and $\chi^{-+}$,
each of which can be evaluated from a Kubo formula, 
\begin{equation}
    \chi^{+-}(\mathbf{r}, \mathbf{r}', t-t') = -\frac{i}{\hbar} \theta(t-t') \left\langle \left[ \hat{n}^+_0(\mathbf{r},t), \hat{n}^-_0(\mathbf{r}',t') \right] \right\rangle_0,
\end{equation}
where
\begin{subequations}
    \begin{equation}
        \hat{n}^+(\mathbf{r})=\hat{\psi}^\dagger_\uparrow(\mathbf{r})\hat{\psi}_\downarrow(\mathbf{r})
    \end{equation}
    and
    \begin{equation}
        \hat{n}^-(\mathbf{r})=\hat{\psi}^\dagger_\downarrow(\mathbf{r})\hat{\psi}_\uparrow(\mathbf{r}),
    \end{equation}
\end{subequations}
while time-dependence is given in the interaction picture and $\pm$ indices can be interchanged to yield $\chi^{-+}$.
Together, the spin-raising and spin-lowering density operators $\hat{n}^+(\mathbf{r})$ and $\hat{n}^-(\mathbf{r})$ generate their own hierarchy of excited states, namely one where $S^z$ is changed by $n\hbar$ with respect to the ground state. 
In the zero temperature limit, the linear susceptibilities $\chi^{+-}$ and $\chi^{-+}$ only involve the $|n|=1$ excitations of this hierarchy. 

To illustrate this, the zero temperature susceptibility is separated into its reactive and dissipative parts, $\chi^{+-}=\chi'^{+-}+i\chi''^{+-}$, each inferrable from the other via a Kramers-Kronig relation. The dissipative part of the response corresponds directly to the imaginary part of the excitation-frequency poles and is expressed here in terms of the scattering function $S^{+-}(\omega)=-\chi''^{+-}(\omega)/\pi$. The scattering function can be further separated into two distinct spectral functions, 
one for the spin-lowering excitations $A^{+-}$ and one for the spin-raising excitations $A^{-+}$ \cite{Skovhus2021,Skovhus2024},
\begin{equation}\label{eq:scattering function}
    S^{+-}(\mathbf{r}, \mathbf{r}', \omega) = A^{+-}(\mathbf{r}, \mathbf{r}', \omega) - A^{-+}(\mathbf{r}', \mathbf{r}, -\omega).
\end{equation}
These spectral functions are essentially weighted joint density of states,
\begin{equation}\label{eq:spectral functions}
    A^{+-}(\mathbf{r}, \mathbf{r}', \omega) = \sum_{\alpha>0} n^+_{0\alpha}(\mathbf{r}) n^-_{\alpha0}(\mathbf{r}')\, \delta\hspace{-1pt}(\hbar\omega - [E_\alpha - E_0]),
\end{equation}
where $\pm$ indices can be interchanged to yield $A^{-+}$, $E_\alpha$ are the eigenenergies of the eigenstates $|\alpha\rangle$ and
\begin{equation}
    n^+_{\alpha\alpha'}(\mathbf{r}) = \langle \alpha| \hat{n}^+(\mathbf{r}) |\alpha'\rangle = \left[n^-_{\alpha'\alpha}(\mathbf{r})\right]^*.
\end{equation}
The dissipative part of the susceptibility is thus composed of peaks at positive frequencies $\omega$ corresponding to excitation energies $E_\alpha-E_0$ of states $|\alpha\rangle$ that lower $S^z$ by $\hbar$ and peaks at negative frequencies $-\omega$ corresponding to excitations that raise $S^z$ by $\hbar$.
Therefore, the aim of theoretical magnon spectroscopy is to calculate the transverse magnetic susceptibility $\chi^{+-}$ and extract the scattering function $S^{+-}$, from which the single magnon excitations of a given system can be characterized, along with all other electronic excitations that change $S^z$ by $\pm\hbar$. In systems with strong spin-orbit coupling the aim is similar, but since $S^z$ no longer is a good quantum number, the poles of the susceptibility $\chi^{\mu\nu}$ may not be fully transverse or longitudinal, but rather a mixture of both.

\subsection{Linear response time-dependent density functional theory}\label{sec:lrtddft theory}

In LR-TDDFT, the four-component susceptibility tensor $\chi^{\mu\nu}$ is computed based on a brute-force evaluation of its Kohn-Sham analogue. Evaluating $\chi^{\mu\nu}_\mathrm{KS}$ is, in principle, straightforward, as it can be constructed based on the single-particle Kohn-Sham eigenvalues $\epsilon_n$ and ground state occupations $f_n$,
\begin{equation}
    \chi^{\mu\nu}_\mathrm{KS}(\mathbf{r}, \mathbf{r}', \omega) = \lim_{\eta\rightarrow 0^+} \sum_{n,m} \left(f_n - f_m\right) \frac{n_{nm}^\mu(\mathbf{r}) n_{mn}^\nu(\mathbf{r}')}{\hbar\omega - (\epsilon_m - \epsilon_n) + i\hbar\eta},
    \label{eq:four-component Kohn-Sham susceptibility}
\end{equation}
with matrix elements $n_{nm}^\mu(\mathbf{r})$ calculated from the single-particle Kohn-Sham orbitals $|\psi_n\rangle$,
\begin{equation}
   n_{nm}^\mu(\mathbf{r}) \equiv \langle \psi_n| \hat{n}^\mu(\mathbf{r}) |\psi_m\rangle.
\end{equation}
Crucially, $\chi^{\mu\nu}_\mathrm{KS}$ is not just the four-component susceptibility tensor of an arbitrary single-particle system. It is the susceptibility of \textit{the} auxiliary system where the time-dependent four-component density $n^\mu(\mathbf{r},t)$ is identical to the density of the physical many-body system.
Whereas the many-body susceptibility yields the linear order response $\delta n^\mu$ at time $t$ and position $\mathbf{r}$ due to an external perturbation at a previous time $t'$ and position $\mathbf{r}'$, 
\begin{equation}
    \chi^{\mu\nu}(\mathbf{r},\mathbf{r}',t-t') = \frac{\delta n^\mu(\mathbf{r}, t)}{\delta W_\mathrm{ext}^\nu(\mathbf{r}',t')},
    \label{eq:chi functional derivative}
\end{equation}
$\chi_\mathrm{KS}^{\mu\nu}$ yields the identical response, but due to the corresponding change in effective single-particle electromagnetic potential, $W_\mathrm{s}^\mu=\delta^{\mu 0} V_\mathrm{nuc} + W_\mathrm{ext}^\mu + W_\mathrm{Hxc}^\mu$:
\begin{equation}
    \chi_\mathrm{KS}^{\mu\nu}(\mathbf{r},\mathbf{r}',t-t') = \frac{\delta n^\mu(\mathbf{r}, t)}{\delta W_\mathrm{s}^\nu(\mathbf{r}',t')}.
    \label{eq:chiks functional derivative}
\end{equation}
The difference is the induced change to the Hartree-exchange-correlation potential $W^\mu_\mathrm{Hxc}$ that fluctuations in the four-component density entail,
\begin{equation}
    K_\mathrm{Hxc}^{\mu\nu}(\mathbf{r},\mathbf{r}',t-t') = \frac{\delta W_\mathrm{Hxc}^\mu(\mathbf{r}, t)}{\delta n^\nu(\mathbf{r}',t')}.
    \label{eq:Kxc functional derivative}
\end{equation}
Accounting for this difference, the many-body susceptibility can be obtained by inverting a formally exact Dyson-like equation \cite{Gross1985},
\begin{widetext}
\begin{equation}
    \chi^{\mu\nu}(\mathbf{r},\mathbf{r}',t-t') = \chi_\mathrm{KS}^{\mu\nu}(\mathbf{r},\mathbf{r}',t-t') + 
    \iint d1\,d2\:\:
    \chi_\mathrm{KS}^{\mu\tau_1}(\mathbf{r},\mathbf{r}_1,t-t_1) K_\mathrm{Hxc}^{\tau_1\tau_2}(\mathbf{r}_1,\mathbf{r}_2,t_1-t_2) \chi^{\tau_2\nu}(\mathbf{r}_2,\mathbf{r}',t_2-t'),
    \label{eq:full dyson}
\end{equation}
\end{widetext}
%
where $\int d1\: \equiv \int d\mathbf{r}_1 dt_1$ and summation over repeated $\tau$ indices is implied. 
%
%

The many-body correlations that the Dyson equation \eqref{eq:full dyson} introduces via $K^{\mu\nu}_\mathrm{Hxc}$ cannot be understood based on the Hartree-exchange-correlation kernel alone. Instead, the kernel always appears under an integral with the Kohn-Sham susceptibility. 
\begin{figure}[tb]
    \centering
    \includegraphics[width=\linewidth]{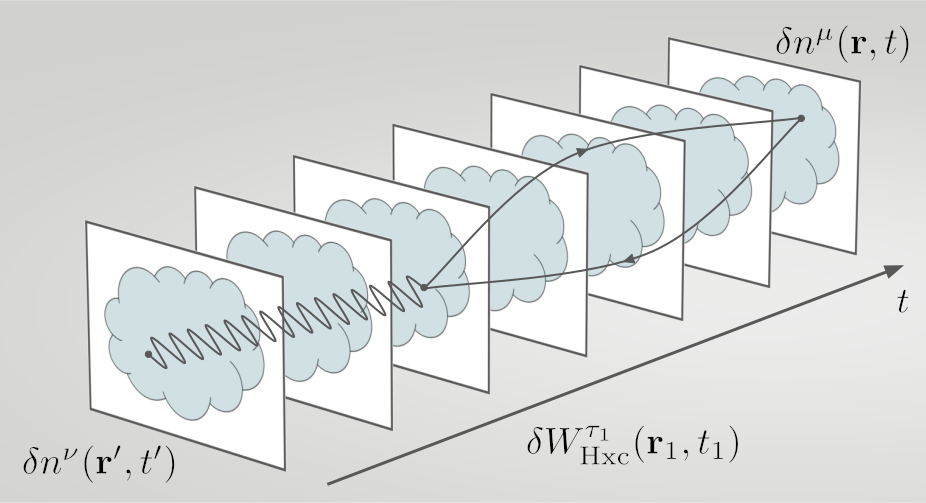}
    \caption{Cartoon of the physics encoded in the self-enhancement function \eqref{eq:self-enhancement function def}. Illustrates how a change (with respect to the ground state) of the four-component electron density (represented by the cloud) in the past, $\delta n^\nu(\mathbf{r}',t')$, invokes fluctuations $\delta n^\mu(\mathbf{r},t)$ in the future via the induced changes to the Hartree-exchange-correlation potential $\delta W_\mathrm{Hxc}^{\tau_1}(\mathbf{r}_1, t_1)$ in the interim, $t' \leq t_1 \leq t$. 
    The wiggly line represents the kernel \eqref{eq:Kxc functional derivative}, while the bubble represents the Kohn-Sham susceptibility \eqref{eq:chiks functional derivative}.
    }
    \label{fig:Self-enhancement function}
\end{figure}
To better illustrate the correlation effects, it is therefore helpful to introduce the notion of a self-enhancement function: 
%
\begin{subequations}\label{eq:self-enhancement function def}
\begin{align}
    \Xi^{\mu\nu}(\mathbf{r},\mathbf{r}',t-t') &\equiv \int d1\: \frac{\delta n^\mu(\mathbf{r}, t)}{\delta W_\mathrm{s}^{\tau}(\mathbf{r}_1,t_1)} \frac{\delta W_\mathrm{Hxc}^{\tau}(\mathbf{r}_1, t_1)}{\delta n^\nu(\mathbf{r}',t')}\\
    &=\int d1\: 
    \chi_\mathrm{KS}^{\mu\tau}(\mathbf{r},\mathbf{r}_1,t-t_1)\notag\\
    &\qquad\qquad\times K_\mathrm{Hxc}^{\tau\nu}(\mathbf{r}_1,\mathbf{r}',t_1-t')
\end{align}
\end{subequations}
%
%
By construction, the self-enhancement function yields the linear order change to the four-component density at time $t$ which is induced by the fluctuations in the Hxc potential that transpire between times $t'$ and $t$ in response to a change in the very same four-component density at a previous time $t'$, 
see Fig. \ref{fig:Self-enhancement function}. Thus, it encodes the electronic system's enhancement and dampening of internal fluctuations, that is, to what extent fluctuations in the past impose fluctuations in the future. 
In fact, if there are frequencies where the self-enhancement function has an eigenvalue of unity, it entails the existence of a fluctuational eigenmode $\delta n^\mu(\mathbf{r}, \omega)$ which the electronic system can sustain on its own via the effective electron-electron interaction. 
The role of the Dyson equation \eqref{eq:full dyson} is to translate such a unity eigenmode into a pole in the many-body susceptibility $\chi^{\mu\nu}$, that is, an excited eigenstate of the many-body system.

\subsection{Bloch susceptibility and collective quasi-particles}\label{sec:bloch susceptibility theory}

In periodic crystals, the many-body susceptibility, Kohn-Sham susceptibility and self-enhancement function are all 
invariant under lattice translations $\chi^{\mu\nu}(\mathbf{r}+\mathbf{R},\mathbf{r}'+\mathbf{R}, \omega)=\chi^{\mu\nu}(\mathbf{r},\mathbf{r}', \omega)$. This invariance implies that the lattice Fourier transform of each quantity,
\begin{equation}
    \chi^{\mu\nu}(\mathbf{r},\mathbf{r}', \mathbf{q}, \omega) = \sum_{\mathbf{R}'} e^{i\mathbf{q}\cdot\mathbf{R}'} \chi^{\mu\nu}(\mathbf{r},\mathbf{r}'+\mathbf{R}', \omega),
\end{equation}
transforms like a Bloch wave,
\begin{equation}
    \chi^{\mu\nu}(\mathbf{r},\mathbf{r}', \mathbf{q}, \omega) = e^{i\mathbf{q}(\mathbf{r}-\mathbf{r}')} \bar{\chi}^{\mu\nu}(\mathbf{r},\mathbf{r}', \mathbf{q}, \omega),
\end{equation}
where the Bloch susceptibility $\bar{\chi}$ is periodic in both spatial entries independently, $\bar{\chi}^{\mu\nu}(\mathbf{r}+\mathbf{R}, \mathbf{r}', \mathbf{q}, \omega) = \bar{\chi}^{\mu\nu}(\mathbf{r},\mathbf{r}'+\mathbf{R}, \mathbf{q}, \omega) = \bar{\chi}^{\mu\nu}(\mathbf{r},\mathbf{r}', \mathbf{q}, \omega)$, see e.g. \cite{SkovhusPhD}. Given some orthonormal basis for the periodic degrees of freedom, see Appendix \ref{app:bloch susceptibilities}, the Dyson equation \eqref{eq:full dyson} becomes a simple matrix equation,
\begin{equation}
    \chi^{\mu\nu}(\mathbf{q},\omega) = \chi_\mathrm{KS}^{\mu\nu}(\mathbf{q},\omega) + \Xi^{\mu\tau}(\mathbf{q},\omega) \chi^{\tau\nu}(\mathbf{q},\omega),
\end{equation}
where $\chi^{\mu\nu}(\mathbf{q},\omega)$ denotes the matrix representation of $\bar{\chi}^{\mu\nu}(\mathbf{r},\mathbf{r}', \mathbf{q}, \omega)$. 
As a result, the many-body susceptibility can be obtained for a single value of $\mathbf{q}$ and $\omega$ at a time, simply by performing a matrix inversion in the periodic basis and spin-components,
\begin{equation}
    \chi^{\mu\nu}(\mathbf{q}, \omega) = \left[1 - \Xi(\mathbf{q}, \omega)\right]^{-1}_{\mu\tau} \chi_\mathrm{KS}^{\tau\nu}(\mathbf{q}, \omega).
    \label{eq:xi dyson periodic systems}
\end{equation}
Cast this way, the role of the $\Xi$ eigenmodes becomes especially clear. For wave vectors $\mathbf{q}$ and frequencies $\omega$ where the single-particle Kohn-Sham susceptibility is pole-less, a pole can still arise in the many-body susceptibility if the self-enhancement function has an eigenmode with unity eigenvalue. The excited states which give rise to such poles in $\chi$ are therefore inherently of a collective nature, thanks also to their interpretation as self-sustained fluctuations in the electron density due to the effective electron-electron interaction of the system. In other words, spectral peaks corresponding to unity eigenvalues of $\Xi$ constitute the \textit{collective excitations} of the electron system, whereas peaks due to poles in $\chi_\mathrm{KS}$ correspond to \textit{single-particle-like} excitations. In the language of quasi-particles, the collective eigenmodes are assigned names such as plasmons or magnons depending on their spin character, with dispersion relations defined from the wave vector dependence of the pole frequency.

\subsection{Transverse magnetic susceptibility in the ALDA}\label{sec:theory ALDA}

Applying the adiabatic local density approximation to collinear systems absent of spin-orbit coupling, the transverse and longitudinal components of the Dyson equation \eqref{eq:full dyson} decouple \cite{Skovhus2021}. The result is a spin-diagonal Dyson equation for the circular coordinate susceptibilities, given here in terms of the Bloch susceptibility,
\begin{equation}
    \chi^{+-}(\mathbf{q},\omega) = \chi_\mathrm{KS}^{+-}(\mathbf{q},\omega) + \Xi^{++}(\mathbf{q},\omega) \chi^{+-}(\mathbf{q},\omega).
    \label{eq:ALDA dyson equation}
\end{equation}
Since the kernel is local in both space and time, the self-enhancement function $\Xi^{++}$ is particularly simple,
\begin{equation}
    \Xi^{++}(\mathbf{r}, \mathbf{r}', t-t') = \chi_\mathrm{KS}^{+-}(\mathbf{r},\mathbf{r}',t-t') f_\mathrm{LDA}^{-+}(\mathbf{r}'),
\end{equation}
where the LDA kernel $f_\mathrm{LDA}^{-+}$ is given by the local ratio of effective Zeeman potential to spin-polarization density:
\begin{equation}
    f_\mathrm{LDA}^{-+}(\mathbf{r}) = f_\mathrm{LDA}^{-+}[n,n^z](\mathbf{r}) = \frac{2 W_\mathrm{xc}^z[n,n^z](\mathbf{r})}{n^z(\mathbf{r})}.
\end{equation}
Importantly, this means that the poles of the self-enhancement function are identical to those of the Kohn-Sham susceptibility. More precisely, introducing the pair density
\begin{equation}
    n_{n\mathbf{k}s,m\mathbf{k}'s'}(\mathbf{r}) = \psi^*_{n\mathbf{k}s}(\mathbf{r}) \psi_{m\mathbf{k}'s'}(\mathbf{r})
    \label{eq:pair density}
\end{equation}
along with the transverse magnetic pair potential 
\begin{equation}
    W^\perp_{n\mathbf{k}s,m\mathbf{k}'s'}(\mathbf{r}) = \psi^*_{n\mathbf{k}s}(\mathbf{r}) \psi_{m\mathbf{k}'s'}(\mathbf{r}) f_\mathrm{LDA}^{-+}(\mathbf{r})
    \label{eq:transverse magnetic pair potential}
\end{equation}
with Kohn-Sham orbitals $\psi_{n\mathbf{k}s}(\mathbf{r})$ normalized to the unit cell, the lattice Fourier transforms of the Kohn-Sham susceptibility and self-enhancement function are given by
\begin{align}
    \chi_\mathrm{KS}^{+-}(\mathbf{r},\mathbf{r}', \mathbf{q}, \omega) = &\lim_{\eta\rightarrow 0^+} \frac{1}{N_k} \sum_\mathbf{k} \sum_{n,m} (f_{n\mathbf{k}\uparrow} - f_{m\mathbf{k}+\mathbf{q}\downarrow})
    \nonumber \\
    &\times \frac{n_{n\mathbf{k}\uparrow,m\mathbf{k}+\mathbf{q}\downarrow}(\mathbf{r}) n_{m\mathbf{k}+\mathbf{q}\downarrow,n\mathbf{k}\uparrow}(\mathbf{r}')}{\hbar\omega - (\epsilon_{m\mathbf{k}+\mathbf{q}\downarrow} - \epsilon_{n\mathbf{k}\uparrow}) + i\hbar\eta}
\end{align}
and
\begin{align}
    \Xi^{++}(\mathbf{r},\mathbf{r}', \mathbf{q}, \omega) = &\lim_{\eta\rightarrow 0^+} \frac{1}{N_k} \sum_\mathbf{k} \sum_{n,m} (f_{n\mathbf{k}\uparrow} - f_{m\mathbf{k}+\mathbf{q}\downarrow})
    \nonumber \\
    &\times \frac{n_{n\mathbf{k}\uparrow,m\mathbf{k}+\mathbf{q}\downarrow}(\mathbf{r}) W^\perp_{m\mathbf{k}+\mathbf{q}\downarrow,n\mathbf{k}\uparrow}(\mathbf{r}')}{\hbar\omega - (\epsilon_{m\mathbf{k}+\mathbf{q}\downarrow} - \epsilon_{n\mathbf{k}\uparrow}) + i\hbar\eta}.
    \label{eq:magnon self-enhancement function}
\end{align}
Thus, both the Kohn-Sham susceptibility and the self-enhancement function exhibit poles exclusively inside the noninteracting Stoner continuum, that is, the continuum of excitations between spin-up and spin-down bands on opposite sides of the Fermi level. Furthermore, collective magnon excitations can only appear outside this continuum according to the Dyson equation \eqref{eq:ALDA dyson equation} under the condition that $\Xi^{++}$ has a unity eigenvalue.

\subsection{Goldstone criterion}

The most fundamental magnon excitations are those of the Goldstone mode. In systems absent of spin-orbit coupling, global spin-rotation symmetry is spontaneously broken by the presence of long-range magnetic order, which---thanks to the Goldstone theorem---entails the existence of a magnon eigenmode with a gapless dispersion. In the language of the self-enhancement function for collinear systems, this means that $\Xi^{++}(\mathbf{q}=\mathbf{0}, \omega=0)$ has a unity eigenvalue.
There exists several starting points to show that this is indeed the case \cite{Katsnelson2004,BuczekPhD,Lounis2011,Rousseau2012}. One is to make explicit use of the spin-rotation symmetry of the Kohn-Sham system, that is, that a rigid rotation of the ground state exchange-correlation field $\mathbf{B}_\mathrm{xc}$ leads to a rigid rotation of the magnetization $\mathbf{m}$ in the static limit. To linear order in the infinitesimal rotation angle $\theta$ \cite{BuczekPhD}, this implies that the Kohn-Sham susceptibility \eqref{eq:chiks functional derivative} and exchange-correlation kernel \eqref{eq:Kxc functional derivative} satisfy
\begin{equation}
    \int d\mathbf{r}'\: \chi_\mathrm{KS}^{+-}(\mathbf{r}, \mathbf{r}', \omega=0) W_\mathrm{xc}^z(\mathbf{r}') = \frac{n^z(\mathbf{r})}{2}
    \label{eq:chiks spin-rotational invariance}
\end{equation}
and
\begin{equation}
    \int d\mathbf{r}'\: \breve{K}_\mathrm{xc}^{-+}(\mathbf{r}, \mathbf{r}', \omega=0) n^z(\mathbf{r}') = 2 W_\mathrm{xc}^z(\mathbf{r}),
    \label{eq:Kxc spin-rotational invariance}
\end{equation}
where the circular components of the kernel follow the notation of Ref. \cite{Skovhus2021} and
\begin{equation}
    \Xi^{++}(\mathbf{r}, \mathbf{r}', \omega) = \int d\mathbf{r}_1\: \chi_\mathrm{KS}^{+-}(\mathbf{r}, \mathbf{r}_1, \omega) \breve{K}_\mathrm{xc}^{-+}(\mathbf{r}_1, \mathbf{r}', \omega).
\end{equation}
In combination, Eqs. \eqref{eq:chiks spin-rotational invariance} and \eqref{eq:Kxc spin-rotational invariance} imply that the eigenvector of the Goldstone magnon mode is the ground state spin-polarization $n^z(\mathbf{r)}$ itself,
\begin{equation}
    \Xi^{++}(\mathbf{q}=\mathbf{0},\omega=0) |n^z\rangle = |n^z\rangle,
    \label{eq:Goldstone criterion}
\end{equation}
why it is also often referred to as the acoustic magnon mode. The Goldstone criterion \eqref{eq:Goldstone criterion} is valid for arbitrary periodic basis representations of the self-enhancement function, but the representability of the ground state spin-polarization within the chosen basis clearly becomes a key convergence parameter.

\section{Implementation}\label{sec:implementation}

Before diving into the details of the implementation, 
it is necessary to emphasize that LR-TDDFT is not usually framed in terms of the self-enhancement function. Instead, preceding implementations express and invert the ALDA Dyson equation \eqref{eq:ALDA dyson equation} using a periodic basis representation of the xc kernel \cite{Buczek2011b,Lounis2011,Rousseau2012,Singh2019,Skovhus2021},
\begin{equation}
    \chi^{+-}(\mathbf{q},\omega) = \chi_\mathrm{KS}^{+-}(\mathbf{q},\omega) + \chi_\mathrm{KS}^{+-}(\mathbf{q},\omega) \breve{K}_\mathrm{xc}^{-+} \chi^{+-}(\mathbf{q},\omega).
    \label{eq:old-fashioned ALDA dyson}
\end{equation}
%
This means that numerical satisfaction of the Goldstone criterion $\eqref{eq:Goldstone criterion}$ hinges on the collective satisfaction of both Eqs. \eqref{eq:chiks spin-rotational invariance} and \eqref{eq:Kxc spin-rotational invariance}. 
That, however, is hard to guarantee in practice, despite the formal triviality of Eq. \eqref{eq:Kxc spin-rotational invariance} for the ALDA, $\breve{K}_\mathrm{xc}^{-+}(\mathbf{r}, \mathbf{r}', \omega) = \delta(\mathbf{r}-\mathbf{r}') f_\mathrm{LDA}^{-+}(\mathbf{r})$. The result is a finite acoustic magnon energy at the $\Gamma$-point, referred to as the Goldstone gap error. 

In order to make progress, we suggest to leverage the locality of the LDA kernel and project the entire self-enhancement function \eqref{eq:magnon self-enhancement function} onto the periodic basis set of choice, calculating the many-body susceptibility from Eq. \eqref{eq:ALDA dyson equation} instead of Eq. \eqref{eq:old-fashioned ALDA dyson}. In this way, satisfaction of the Goldstone criterion \eqref{eq:Goldstone criterion} is no longer limited by the accuracy of the $\chi_\mathrm{KS}^{+-}(\mathbf{q},\omega) \breve{K}_\mathrm{xc}^{-+}$ matrix product, which normally is calculated in an incomplete periodic basis representation. Instead, the product is effectively carried out in real space when evaluating the transverse magnetic pair potential \eqref{eq:transverse magnetic pair potential}.

In this section, we investigate this strategy in the context of the open-source GPAW electronic structure code \cite{Mortensen2005,Enkovaara2010,Mortensen2024}. 
The GPAW linear response code \cite{Yan2011,Skovhus2021} uses a plane-wave basis and already includes functionality to calculate the transverse magnetic susceptibility through Eq. \eqref{eq:old-fashioned ALDA dyson}.
The preexisting code has successfully been used to study the magnon spectrum of both itinerant ferro- and antiferromagnets \cite{Skovhus2021,Skovhus2022a,Skovhus2022b}. However, it fails to satisfy the Goldstone criterion \eqref{eq:Goldstone criterion} numerically within a truncated plane-wave basis \cite{Skovhus2021}. 
Here, we present a novel PAW implementation that allows us to evaluate the self-enhancement function \eqref{eq:magnon self-enhancement function} literally and without numerical inconsistency. 
Inverting the Dyson Eq. \eqref{eq:ALDA dyson equation} in a plane-wave basis, we investigate the gap error 
in the limit of a complete basis for itinerant ferromagnets Fe, Ni and Co, and in Appendix \ref{app:magnon dispersion convergence} we present the band and basis set convergence of the magnon dispersion.  

\subsection{Plane-wave representation}

Using plane waves $e^{i\mathbf{G}\cdot\mathbf{r}}/\sqrt{\Omega_\mathrm{cell}}$ as basis functions for the periodic degrees of freedom, the matrix representation of $\chi^{+-}(\mathbf{q},\omega)$ is given by a Fourier transform,
\begin{align}
    \chi^{+-}_{\mathbf{G}\mathbf{G}'}(\mathbf{q},\omega) 
    = &\frac{1}{\Omega_\mathrm{cell}} \iint_{\Omega_\mathrm{cell}} d\mathbf{r}d\mathbf{r}'\: e^{-i(\mathbf{G}+\mathbf{q})\cdot\mathbf{r}}
    \nonumber \\
    &\times
    \chi^{+-}(\mathbf{r},\mathbf{r}',\mathbf{q},\omega) e^{i(\mathbf{G}'+\mathbf{q})\cdot\mathbf{r}'}.
\end{align}
The plane-wave susceptibility $\chi^{+-}_{\mathbf{G}\mathbf{G}'}(\mathbf{q},\omega)$ is obtained from the Fourier transformed Kohn-Sham susceptibility 
\begin{align}
    \chi_\mathrm{KS,\mathbf{G}\mathbf{G}'}^{+-}&(\mathbf{q}, \omega) = 
    \lim_{\eta\rightarrow 0^+} \frac{1}{\Omega} \sum_\mathbf{k} \sum_{n,m} (f_{n\mathbf{k}\uparrow} - f_{m\mathbf{k}+\mathbf{q}\downarrow})
    \nonumber \\
    &\times \frac{n_{n\mathbf{k}\uparrow,m\mathbf{k}+\mathbf{q}\downarrow}(\mathbf{G}+\mathbf{q}) n_{m\mathbf{k}+\mathbf{q}\downarrow,n\mathbf{k}\uparrow}(-\mathbf{G}'-\mathbf{q})}{\hbar\omega - (\epsilon_{m\mathbf{k}+\mathbf{q}\downarrow} - \epsilon_{n\mathbf{k}\uparrow}) + i\hbar\eta}
    \label{eq:plane-wave Kohn-Sham susceptibility}
\end{align}
and self-enhancement function
\begin{align}
    \Xi^{++}_{\mathbf{G}\mathbf{G}'}&(\mathbf{q}, \omega) = 
    \lim_{\eta\rightarrow 0^+} \frac{1}{\Omega} \sum_\mathbf{k} \sum_{n,m} (f_{n\mathbf{k}\uparrow} - f_{m\mathbf{k}+\mathbf{q}\downarrow})
    \nonumber \\
    &\times
    \frac{n_{n\mathbf{k}\uparrow,m\mathbf{k}+\mathbf{q}\downarrow}(\mathbf{G}+\mathbf{q}) W^\perp_{m\mathbf{k}+\mathbf{q}\downarrow,n\mathbf{k}\uparrow}(-\mathbf{G}'-\mathbf{q})}{\hbar\omega - (\epsilon_{m\mathbf{k}+\mathbf{q}\downarrow} - \epsilon_{n\mathbf{k}\uparrow}) + i\hbar\eta},
    \label{eq:plane-wave self-enhancement function}
\end{align}
each constructed from the Fourier transformed matrix elements
\begin{equation}
    n_{n\mathbf{k}s,m\mathbf{k}'s'}(\mathbf{G}+\mathbf{q}) = \int_{\Omega_\mathrm{cell}}d\mathbf{r}\: e^{-i(\mathbf{G}+\mathbf{q})\cdot\mathbf{r}} n_{n\mathbf{k}s,m\mathbf{k}'s'}(\mathbf{r})
    \label{eq:plane-wave pair density}
\end{equation}
and
\begin{equation}
    W^\perp_{n\mathbf{k}s,m\mathbf{k}'s'}(\mathbf{G}+\mathbf{q}) = \int_{\Omega_\mathrm{cell}}d\mathbf{r}\: e^{-i(\mathbf{G}+\mathbf{q})\cdot\mathbf{r}} W^\perp_{n\mathbf{k}s,m\mathbf{k}'s'}(\mathbf{r}).
    \label{eq:plane-wave pair potential}
\end{equation}
Since the GPAW linear response code already supplies functionality to compute plane-wave pair densities \eqref{eq:plane-wave pair density}, all which is needed is a generalization of this code to plane-wave pair potentials of the form \eqref{eq:plane-wave pair potential}.
With the plane-wave matrix elements in hand, it is then straightforward to compute the self-enhancement function \eqref{eq:plane-wave self-enhancement function} using the same underlying code as is used to compute the Kohn-Sham susceptibility \eqref{eq:plane-wave Kohn-Sham susceptibility}, upon which the Dyson Eq. \eqref{eq:ALDA dyson equation} can be inverted numerically.

\subsection{PAW implementation}

In the PAW method \cite{Blochl1994}, matrix elements of the form
\begin{equation}
    W_{n\mathbf{k}s,m\mathbf{k}'s'}(\mathbf{G}+\mathbf{q}) = \langle \psi_{n\mathbf{k}s}| e^{-i(\mathbf{G}+\mathbf{q})\cdot \mathbf{r}} f(\mathbf{r}) |\psi_{m\mathbf{k}'s'}\rangle_{\Omega_\mathrm{cell}}
    \label{eq:generalized pair potential}
\end{equation}
can be evaluated efficiently using smooth representations of the Kohn-Sham orbitals, provided that the functional $f(\mathbf{r})=f[n](\mathbf{r})$ itself is sufficiently smooth. The evaluation can, without loss of generality, be split into a pseudo contribution and a PAW correction,
\begin{align}
    W_{n\mathbf{k}s,m\mathbf{k}'s'}(\mathbf{G}+\mathbf{q}) = 
    &\tilde{W}_{n\mathbf{k}s,m\mathbf{k}'s'}(\mathbf{G}+\mathbf{q}) 
    \nonumber \\
    &+ \Delta W_{n\mathbf{k}s,m\mathbf{k}'s'}(\mathbf{G}+\mathbf{q}),
\end{align}
where the pseudo contribution is computed as a fast Fourier transform (FFT) of the smooth pseudo waves $\tilde{\psi}_{n\mathbf{k}s}$ and the functional $f$,
\begin{equation}
    \tilde{W}_{n\mathbf{k}s,m\mathbf{k}'s'}(\mathbf{G}+\mathbf{q}) = \mathcal{F}_\mathbf{G}[ e^{-i\mathbf{q}\cdot\mathbf{r}} \tilde{\psi}^*_{n\mathbf{k}s}(\mathbf{r}) \tilde{\psi}_{m\mathbf{k}'s'}(\mathbf{r}) f(\mathbf{r}) ].
\end{equation}
In order to guarantee the accuracy of the FFT, the integrand $e^{-i\mathbf{q}\cdot\mathbf{r}} \tilde{\psi}^*_{n\mathbf{k}s}(\mathbf{r}) \tilde{\psi}_{m\mathbf{k}+\mathbf{q}s'}(\mathbf{r}) f(\mathbf{r})$ needs to be smooth on the real-space grid of the transform. This is well satisfied for the xc kernel in the local density approximation, thanks to its $n^{-2/3}$ scaling with the electron density. Inside the atomic augmentation spheres---where the electron density varies rapidly---the kernel smoothly vanishes making the integrand smooth in the pseudo-wave construction. Outside the augmentation spheres---where the pseudo waves equal the all-electron wave functions and the electron density gradually vanishes---the integrand scales roughly as $n^{1/3}$, guaranteeing smoothness in turn. The latter case is especially important in the presence of vacuum. The divergence of the kernel in vacuum makes its Fourier representation numerically ill behaved and inversion of the Dyson equation \eqref{eq:old-fashioned ALDA dyson} becomes fruitless in a plane-wave basis. In contrast, the plane-wave representation of the self-enhancement function \eqref{eq:plane-wave self-enhancement function} remains well behaved, thanks to the exponential decay of both the pair density \eqref{eq:pair density} and pair potential \eqref{eq:transverse magnetic pair potential} in vacuum. This has allowed us to use Eq. \eqref{eq:ALDA dyson equation} and the implementation presented here to compute the magnon spectrum of e.g. the Fe$_3$GeTe$_2$ monolayer \cite{Skovhus2024}.

Calculating the pseudo contribution with a FFT leaves out only the PAW correction,
\begin{align}
    \Delta W_{n\mathbf{k}s,m\mathbf{k}'s'}(\mathbf{G}+\mathbf{q}) = &\sum_a \sum_{i,i'} \Delta W_{ii'}^a(\mathbf{G}+\mathbf{q}) 
    \nonumber \\
    &\times \langle \tilde{\psi}_{n\mathbf{k}s} | \tilde{p}_i^a \rangle \langle \tilde{p}_{i'}^a | \tilde{\psi}_{m\mathbf{k}'s'} \rangle.
    \label{eq:PAW correction}
\end{align}
The PAW method uses smooth projector functions $|\tilde{p}_i^a\rangle$ to map the pseudo waves from atom-centered smooth partial waves $\tilde{\phi}_i^a$ onto all-electron partial waves $\phi_i^a$ inside each atomic augmentation sphere $a$. The PAW projections $\langle \tilde{p}_i^a | \tilde{\psi}_{n\mathbf{k}s} \rangle$ for the correction \eqref{eq:PAW correction} can be directly extracted from the underlying DFT ground state calculation, meaning that the main new quantity to calculate is the PAW correction matrices
\begin{align}
    \Delta W_{ii'}^a(\mathbf{K}) = &\int_{\Omega_\mathrm{cell}}d\mathbf{r}\: e^{-i\mathbf{K} \cdot\mathbf{r}}
    \big[\phi_i^{a*}(\mathbf{r}-\mathbf{R}_a)\phi_{i'}^a(\mathbf{r}-\mathbf{R}_a) 
    \nonumber \\
    &- \tilde{\phi}_i^{a*}(\mathbf{r}-\mathbf{R}_a)\tilde{\phi}_{i'}^a(\mathbf{r}-\mathbf{R}_a)\big] f(\mathbf{r}).
    \label{eq:PAW correction matrix}
\end{align}
In GPAW, the partial waves are all defined in terms of real spherical harmonics and real radial functions, $\phi_i^a(\mathbf{r})=Y_{l_i}^{m_i}(\hat{\mathbf{r}})\phi_i^a(r)$. To evaluate the correction matrices \eqref{eq:PAW correction matrix}, we therefore make a real spherical harmonics expansion of the functional $f(\mathbf{r})$ inside each of the augmentation spheres, analogous to how the xc field is expanded in order to calculate exchange couplings \cite{Skovhus2025}. 
%
%
Similarly, also the plane waves are written as an expansion in real spherical harmonics and spherical Bessel functions $j_l(x)$ (see e.g. GPAW's plane-wave pair density implementation \cite{Skovhus2021}). Finally, the correction matrices can be computed from tabulated integrals over four spherical harmonics $G_{l_1l_2l_3l_4}^{m_1m_2m_3m_4}$,
\begin{align}
    \Delta W_{ii'}^a(\mathbf{K}) = &4\pi e^{-i\mathbf{K}\cdot\mathbf{R}_a} \sum_l \sum_{m=-l}^l (-i)^l Y_l^m(\hat{\mathbf{K}}) 
    \nonumber \\
    &\times \sum_{l'} \sum_{m'=-l'}^{l'} G_{l_il_{i'}l'l}^{m_im_{i'}m'm} \Delta W_{ii'}^{a,ll'm'}(|\mathbf{K}|),
\end{align}
with radial integrals,
\begin{align}
    \Delta W_{ii'}^{a,ll'm'}(K) = &\int_0^\infty r^2 dr\: j_l(Kr) \big[\phi_i^a(r)\phi_{i'}^a(r) 
    \nonumber \\
    &- \tilde{\phi}_i^a(r)\tilde{\phi}_{i'}^a(r)
    \big] f_{l'}^{m'}(r),
\end{align}
performed on the dense nonlinear radial grid of GPAW's augmentation spheres.



\subsection{Elimination of the gap error}\label{sec:gap error convergence}

\begin{figure*}[tb]
    \centering
    \includegraphics[scale=1.0]{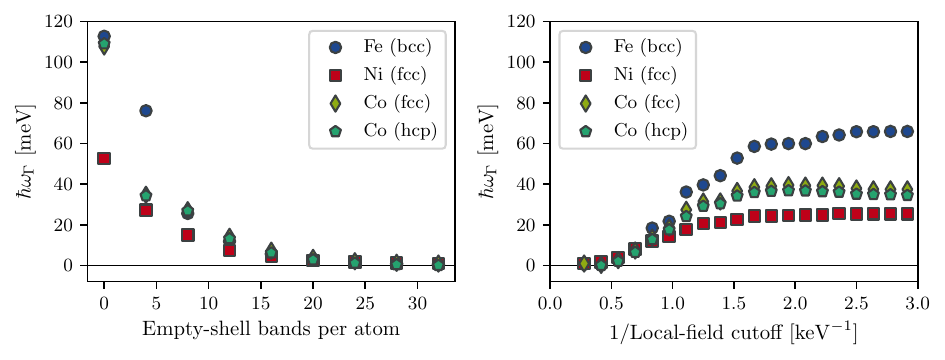}
    \caption{Goldstone gap error in Fe, Ni and Co as a function of the number of empty-shell bands per atom and inverse local-field cutoff. When varying the number of bands (left), the plane-wave cutoff is fixed to 3.6 keV for bcc and fcc structures and 2.4 keV for hcp-Co. When varying the plane-wave cutoff (right), the number of empty-shell bands per atom is fixed to 32.}
    \label{fig:gap_error_convergence}
\end{figure*}
The implementation presented above involves three main convergence parameters for the calculation of the transverse magnetic susceptibility $\chi^{+-}(\mathbf{q},\omega)$: the $k$-point sampling, the number of bands and the plane-wave cutoff. 
The $k$-point convergence has been investigated in detail elsewhere \cite{Skovhus2021}, and here we will simply note that the gap error is \textit{much} less sensitive to the $k$-point sampling than the magnon dispersion at finite wave vectors, thanks to the gapped Stoner continuum at the $\Gamma$-point. 
In Fig. \ref{fig:gap_error_convergence}, we present the convergence of the Goldstone gap error in Fe, Ni and Co as a function of the number of empty-shell bands 
included in Eqs. \eqref{eq:plane-wave Kohn-Sham susceptibility} and \eqref{eq:plane-wave self-enhancement function} as well as the plane-wave cutoff used for the inversion of the Dyson Eq. \eqref{eq:ALDA dyson equation}. The gap error $\hbar\omega_\Gamma$ is itself extracted as the spectral maximum of the majority spectral function \eqref{eq:spectral functions} projected along the Goldstone mode $|n^z\rangle$. For all four crystals, we can effectively eliminate the gap error (to less than 1 meV) provided that a sufficient plane-wave cutoff and number of bands is used. 
To our knowledge, this is the first LR-TDDFT implementation where full gap error elimination can be achieved from raw numerical convergence. We attribute this to the fact that the effect of the local xc kernel is calculated in real space and not in a truncated periodic basis representation. In this way, numerical inconsistencies between Eqs. \eqref{eq:chiks spin-rotational invariance} and \eqref{eq:Kxc spin-rotational invariance} are circumvented and satisfaction of the Goldstone criterion \eqref{eq:Goldstone criterion} is reduced to two key conditions: $|n^z\rangle$ should be an eigenfunction of $\Xi^{++}(\mathbf{q}=\mathbf{0},\omega=0)$ and the eigenvalue should be 1. The first condition is the easiest to satisfy. With enough bands included in Eq. \eqref{eq:plane-wave self-enhancement function}, $\Xi^{++}(\mathbf{q}=\mathbf{0},\omega=0)$ has an eigenfunction $|u_0\rangle$ which is practically identical to the spin-polarization. That is, the absolute error of the (normalized) overlap $|1 - \langle n^z|u_0\rangle|< 5\times10^{-5}$ for all four crystals, regardless of the size of the plane-wave basis used to evaluate the overlap in. Converging the eigenvalue to unity is much harder. 
In principle, one needs to perfectly resolve the full ground state spin-polarization $n^z(\mathbf{r})$ within the periodic basis representation of $\Xi^{++}(\mathbf{q},\omega)$, why several keVs of plane-waves are necessary to eliminate the gap error in Fig. \ref{fig:gap_error_convergence}. 
However, hope is not lost for the LR-TDDFT methodology. Instead of trying to represent the spin-polarization with plane waves, it seems much more promising to simply include $n^z(\mathbf{r})$ as an explicit basis function for $\Xi^{++}(\mathbf{q},\omega)$. In this way, basis set convergence should become trivial for the gap error, and the main convergence parameter would be the number of bands, which based on the results in Fig. \ref{fig:gap_error_convergence} is realistic to converge by brute force for a much wider range of materials.


\section{Transverse magnetic excitations in Fe, Ni and Co}\label{sec:transition metals results}

In literature, there exist plenty \textit{ab initio} accounts of the (A)LDA magnon dispersion in Fe, Ni and Co, all with qualitatively similar results and quantitative disagreements limited to the magnon dispersion inside the Stoner continuum \cite{Savrasov1998,Karlsson2000,SasIoglu2010,Muller2016,Friedrich2020,Buczek2011b,Rousseau2012,Cao2017,Singh2019,Tancogne-Dejean2020,Skovhus2021,Liu2023}. In the discussion below, we will therefore focus more on the physical nature of individual spectral peaks and simply note in passing that the calculated dispersions align well with preexisting trends in literature. Focusing on physical interpretation, we leverage the plane-wave basis of the presented implementation to analyze the eigenmodes of the transverse magnetic susceptibility and discuss in detail the origin of its spectral peaks from the point-of-view of the self-enhancement function.


\subsection{Mode decomposition}\label{sec:mode decomposition}

\begin{figure*}[tb]
    \centering
    \includegraphics[scale=1.0]{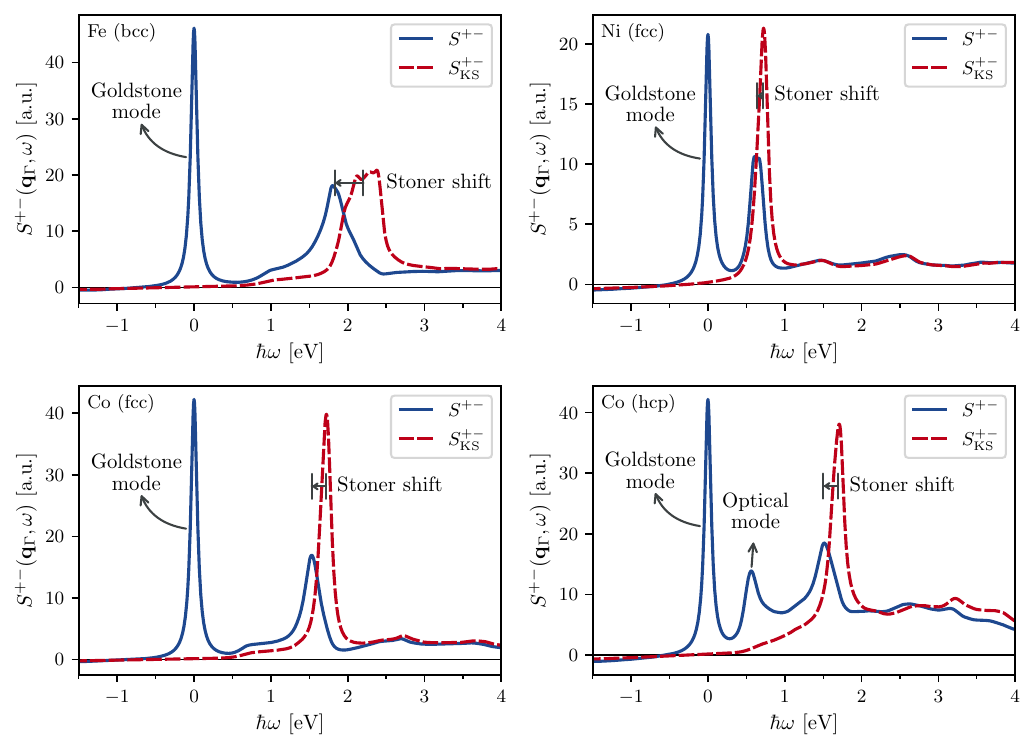}
    \caption{Total scattering function [trace of Eq. \eqref{eq:bloch scattering function}] of Fe, Ni and Co at zero momentum transfer, plotted alongside the total scattering of the single-particle Kohn-Sham system. The Stoner shift indicates the redshift of the Stoner pair gap, using values from Tab. \ref{tab:stoner shift}.}
    \label{fig:total_scattering}
\end{figure*}
Having calculated the transverse magnetic susceptibility, the spectrum of transverse magnetic excitations may be analyzed in terms of the scattering function $S^{+-}$ and its constituents, the spectral functions for spin-lowering and spin-raising excitations $A^{+-}$ and $A^{-+}$, see Eqs. \eqref{eq:scattering function} and \eqref{eq:spectral functions}. For each wave vector $\mathbf{q}$, the scattering function is extracted by taking the anti-Hermitian part (in plane-wave components) of the Bloch susceptibility,
\begin{equation}\label{eq:bloch scattering function}
    S^{+-}(\mathbf{q},\omega) = -\frac{1}{2\pi i}\left[\chi^{+-}(\mathbf{q},\omega) - \chi^{+-}(\mathbf{q},\omega)^\dagger\right],
\end{equation}
from which also the spectral functions can be directly deduced. Namely, the spectral functions are positive-semidefinite, and finite valued only at nonnegative frequencies, 
see also Appendix \ref{app:bloch susceptibilities}. 
Eigendecomposing the scattering function,
\begin{equation}
    \label{eq:bloch scattering function eigendecomposition}
    S^{+-}(\mathbf{q},\omega)|v_n(\mathbf{q},\omega)\rangle=s^{+-}_n(\mathbf{q},\omega)|v_n(\mathbf{q},\omega)\rangle,
\end{equation}
one may thus construct the majority (spin-lowering) spectral function from the positive eigenvalues,
\begin{equation}\label{eq:bloch spectral function reconstructed}
    A^{+-}(\mathbf{q},\omega) = \sum_{n}^{s^{+-}_n(\mathbf{q},\omega)>0} s^{+-}_n(\mathbf{q},\omega) |v_n(\mathbf{q},\omega)\rangle \langle v_n(\mathbf{q},\omega)|.
\end{equation}
%
Strictly speaking, the separation \eqref{eq:bloch spectral function reconstructed} is formally exact only for real frequencies $\omega$, but it works well in practice also when including a finite frequency broadening \cite{Skovhus2024}.

\subsubsection{Total scattering function}

Taking the trace of Eq. \eqref{eq:bloch scattering function}, one obtains a total scattering function for all the charge-neutral many-body excitations that lower/raise (at positive/negative $\omega$) $S^z$ by $\hbar$ and change the crystal momentum by $\hbar\mathbf{q}$ with respect to the ground state. In Fig. \ref{fig:total_scattering} we present the calculated scattering function trace in Fe, Ni and Co at the $\Gamma$-point along with its Kohn-Sham analogue. Comparing the two scattering functions for all four materials, the physical interpretation of each main feature is pretty clear. At positive frequencies, $S_\mathrm{KS}^{+-}$ quantifies the spectral weight of band excitations between pairs of occupied majority states and unoccupied minority states. At the $\Gamma$-point, $S_\mathrm{KS}^{+-}$ is therefore dominated by a single peak corresponding to the exchange splitting of the 3$d$ states. The many-body correlation effects of the Dyson Eq. \eqref{eq:ALDA dyson equation} redshifts the Stoner pair excitations and diminishes their overall weight in $S^{+-}(\mathbf{q},\omega)$, producing instead a new set of collective magnon peaks which are self-sustained in the effective electron-electron interaction. For bcc and fcc structures, a single magnon mode emerges, namely the Goldstone mode peaked at $\omega=0$, but for Co (hcp), which has two magnetic atoms in its unit cell, the Goldstone mode is accompanied by an optical mode with a finite peak frequency.

\subsubsection{Collective and single-particle modes}

\begin{figure*}[tb]
    \centering
    \includegraphics[scale=1.0]{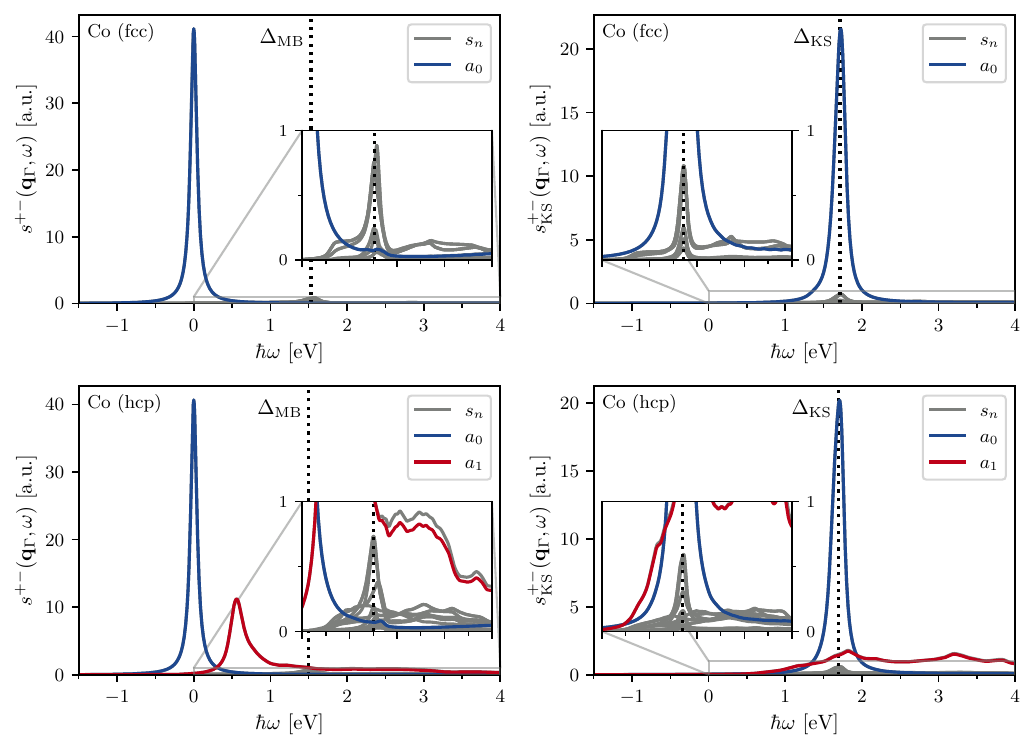}
    \caption{Eigenvalues $s^{+-}_n$ (grey) of the scattering function \eqref{eq:bloch scattering function eigendecomposition} in fcc-Co (top) and hcp-Co (bottom) at zero momentum transfer for the many-body system (left) and Kohn-Sham system (right). For fcc/hcp Co, the 12/24 largest eigenvalues are plotted. In color, the projection \eqref{eq:majority magnon lineshape} of the majority spectral function \eqref{eq:bloch spectral function reconstructed} onto the extracted collective eigenmodes $|v_n(\mathbf{q}_\Gamma)\rangle$ is shown, lying virtually on top of the corresponding eigenvalue for a wide range of frequencies. The effective (Kohn-Sham) exchange splitting $\Delta_\mathrm{KS}$ and (many-body) Stoner pair gap $\Delta_\mathrm{MB}$---see Tab. \ref{tab:stoner shift}---are shown as vertical dotted lines.}
    \label{fig:Co_eigendecomposition}
\end{figure*}
The separation of collective magnon quasi-particles from single-particle Stoner pairs becomes even more clear when looking at the individual eigenvalues of the scattering functions. In Fig. \ref{fig:Co_eigendecomposition}, we present the many-body and Kohn-Sham eigenvalues for Co (fcc and hcp). A corresponding figure for Fe (bcc) and Ni (fcc) is provided in Appendix \ref{app:mode decomposition} along with practical details concerning the storage of plane-wave components. Out of all the eigenvalues in the plane-wave representation, the magnon peaks of Fig. \ref{fig:total_scattering} exist entirely due to a single (collective) eigenmode of $S^{+-}(\mathbf{q},\omega)$ each, with spectral weights order(s) of magnitude larger than the remaining (single-particle) modes. The spectral weight of the single-particle modes is instead restricted to the Stoner continuum (the self-enhancement function never reaches unity for the single-particle modes), with multiple eigenvalues peaking around the many-body Stoner pair gap. 

Not only does the eigendecomposition \eqref{eq:bloch scattering function eigendecomposition} separate the collective magnon modes from the single-particle Stoner modes, it also turns out that the $N$ collective mode vectors ($N=1$ for fcc and bcc, $N=2$ for hcp) are more or less frequency independent, $|v_n(\mathbf{q},\omega)\rangle\simeq |v_n(\mathbf{q})\rangle$. This allows us to define and calculate majority magnon lineshapes,
\begin{equation}\label{eq:majority magnon lineshape}
    a^{+-}_n(\mathbf{q},\omega)=\langle v_n(\mathbf{q})|A^{+-}(\mathbf{q},\omega)|v_n(\mathbf{q})\rangle,
\end{equation}
with mode vectors $|v_n(\mathbf{q})\rangle$ extracted at an appropriately chosen extraction frequency $\omega^\mathrm{e}_\mathbf{q}$.
In this work, $\omega^\mathrm{e}_\mathbf{q}$ is chosen to maximize the minimal many-body eigenvalue difference 
\begin{equation}
    \omega^\mathrm{e}_\mathbf{q}=\max_\omega\left\{\min_{n=0}^{N-1}\left[s^{+-}_n(\mathbf{q},\omega) - s^{+-}_{n+1}(\mathbf{q},\omega)\right]\right\}, 
\end{equation}
with index $n$ sorting the eigenvalues $s^{+-}_n$ in descending order,
thus ensuring maximal separation of the $N+1$ largest eigenvalues at the extraction frequency. However, the presented results are not sensitive to this choice. As long as the collective and single-particle eigenmodes are clearly separated at $\omega^\mathrm{e}_\mathbf{q}$, variations in eigenvectors and lineshapes remain negligible. 
In addition to the bare eigenvalues, we present in Fig. \ref{fig:Co_eigendecomposition} also the many-body magnon lineshape \eqref{eq:majority magnon lineshape} and its Kohn-Sham analogue (projecting $A^{+-}_\mathrm{KS}$ onto the collective mode vectors $|v_n(\mathbf{q})\rangle$).
In both cases, the extracted lineshapes are largely indistinguishable from the corresponding set of scattering function eigenvalues, not only at the extraction frequency itself, but across the plotted frequency range. 
This means that the collective mode vectors $|v_n(\mathbf{q})\rangle$ constitute eigenmodes to the single-particle Kohn-Sham susceptibility as well as the many-body susceptibility. Thus, the primary role of the Dyson equation \eqref{eq:ALDA dyson equation} is to move the spectral weight of the collective eigenmodes---which is peaked at the effective exchange splitting $\Delta_\mathrm{KS}$ in $\chi_\mathrm{KS}^{+-}(\mathbf{q}_\Gamma,\omega)$---onto the magnon resonance(s), and not to rotate the eigenmodes themselves. This is possible because the magnon mode vectors also comprise eigenmodes to the self-enhancement function $\Xi^{++}(\mathbf{q},\omega)$. 
\begin{table}[tb]
    \centering
    \bgroup
    \def\arraystretch{1.4}
    \begin{tabular}{c c|c|c|c|c} 
         & &
        Fe (bcc) & Ni (fcc) & Co (fcc) & Co (hcp) \\
        \hline
        \multirow{2}{.25cm}{\rotatebox[origin=c]{90}{$n=0$}} & $S_\mathrm{KS}^{+-}$ &
        $8.8\times 10^{-4}$ & $7.4\times 10^{-4}$ & $6.9\times 10^{-4}$ & $7.6\times 10^{-4}$ \\
        & $\Xi^{++}$ &
        $6.5\times 10^{-3}$ & $10.2\times 10^{-4}$ & $7.2\times 10^{-4}$ & $9.1\times 10^{-4}$ \\
        \hline
        \multirow{2}{.25cm}{\rotatebox[origin=c]{90}{$n=1$}} & $S_\mathrm{KS}^{+-}$ & 
        & & & $18.2\times 10^{-3}$ \\
        & $\Xi^{++}$ & 
        & & & $16.5\times 10^{-3}$ \\
    \end{tabular}
    \egroup
    \caption{Normalized mode deviation $1 - |\langle u_n(\mathbf{q}, \omega)|v_n(\mathbf{q})\rangle|$ of the dominating eigenvectors $|u_n(\mathbf{q},\omega)\rangle$ of the Kohn-Sham scattering function $S_\mathrm{KS}^{+-}(\mathbf{q},\omega)$ and self-enhancement function $\Xi^{++}(\mathbf{q},\omega)$, evaluated at the $\Gamma$-point and $\hbar\omega=\Delta_\mathrm{KS}$.}
    \label{tab:mode_overlaps}
\end{table}
In Tab. \ref{tab:mode_overlaps} we compare the extracted magnon mode vectors to the eigenvectors of the Kohn-Sham susceptibility and self-enhancement function at $\mathbf{q}_\Gamma$ and $\hbar\omega=\Delta_\mathrm{KS}$. These are all practically identical (and equal to the spin-polarization $|n^z\rangle$ for the Goldstone mode), meaning that the Dyson equation \eqref{eq:ALDA dyson equation} can be written on a block diagonal form, where each collective mode hardly couples to the remaining degrees of freedom. This should, once again, motivate future implementations to explicitly include the spin-polarization $|n^z\rangle$ as a basis function, thus eliminating the truncated basis component of the gap error, see also Sec. \ref{sec:gap error convergence}.

\subsection{Stoner spectra}

\begin{figure*}[tb]
    \centering
    \includegraphics[scale=1.0]{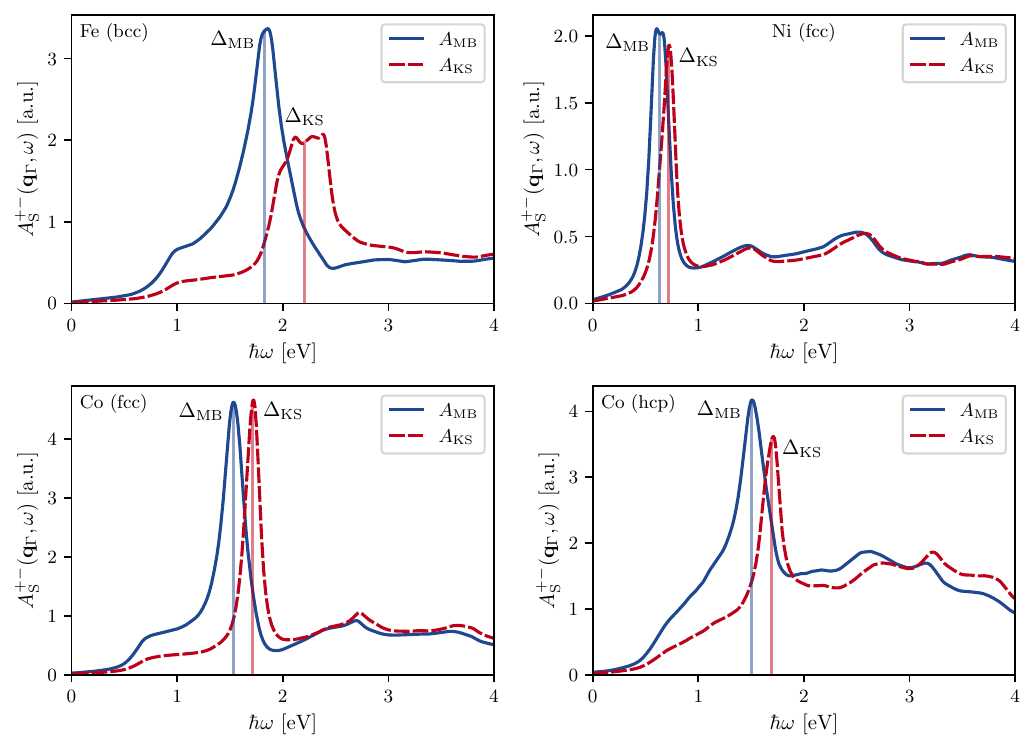}
    \caption{Stoner spectrum [trace of Eq. \eqref{eq:stoner continuum spectral function}] of Fe, Ni and Co at zero momentum transfer, plotted alongside the corresponding Stoner spectrum of the noninteracting Kohn-Sham system. The vertical colored lines indicate the (many-body) Stoner pair gap $\Delta_\mathrm{MB}$ and the effective (Kohn-Sham) exchange splitting $\Delta_\mathrm{KS}$, see Tab. \ref{tab:stoner shift} and Eq. \eqref{eq:stoner pair gap}.}
    \label{fig:stoner_spectrum}
\end{figure*}
From the eigendecomposition of the many-body scattering function $S^{+-}(\mathbf{q},\omega)$ analyzed above, a clear picture has emerged for the transverse magnetic excitations in (idealized) real materials. Out of the infinite degrees of freedom, there exist $N$ collective modes of excitation (magnon modes), where $N$ for elemental Fe, Ni and Co corresponds to the number of (magnetic) atoms. The remaining degrees of freedom contribute instead (in some cases infinitesimally) to the many-body Stoner continuum of electron-hole pair excitations. 
Accordingly, one may define a spectral function for the majority Stoner continuum, simply by subtracting the magnon mode contributions from the spectral function,
\begin{equation}\label{eq:stoner continuum spectral function}
    A^{+-}_\mathrm{S}(\mathbf{q},\omega) \equiv A^{+-}(\mathbf{q},\omega) - \sum_{n=0}^{N-1} a^{+-}_n(\mathbf{q},\omega) |v_n(\mathbf{q})\rangle\langle v_n(\mathbf{q})|.
\end{equation}
Since the collective magnon modes to a good approximation are eigenmodes of the Kohn-Sham scattering function as well, we may similarly subtract $a^{+-}_{\mathrm{KS},n}(\mathbf{q},\omega)=\langle v_n(\mathbf{q})|A^{+-}_\mathrm{KS}(\mathbf{q},\omega)|v_n(\mathbf{q})\rangle$ from $A^{+-}_\mathrm{KS}(\mathbf{q},\omega)$ to define a spectral function for the noninteracting majority Stoner continuum (in the ideal case the entire weight of $a^{+-}_{\mathrm{KS},n}(\mathbf{q},\omega)$ is redistributed to the magnon resonance by the Dyson Eq. \eqref{eq:ALDA dyson equation}).

Based on this definition, we present in Fig. \ref{fig:stoner_spectrum} the many-body and noninteracting Stoner spectra of Fe, Ni and Co in the long wavelength limit. Here, one can clearly observe how the electronic correlations governed by the self-enhancement function $\Xi^{++}(\mathbf{q},\omega)$ redshift the entire spectral weight of the Stoner spectrum along with all its peaks and shoulders. The shift is a redshift due to the bonding nature of the exchange interaction. At the $\Gamma$-point, the noninteracting Stoner spectrum is dominated by a single peak at the exchange splitting of the 3$d$-states (here denoted $\Delta_\mathrm{KS}$). The corresponding peak in the many-body Stoner spectrum may therefore be viewed as a form of many-body Stoner pair gap (denoted $\Delta_\mathrm{MB}$). It quantifies the energy associated with Stoner pair excitations between well separated 3$d$-states of the same orbital character. However, it is not a fundamental gap. Instead, Stoner pair excitations also exist below $\Delta_\mathrm{MB}$ (see e.g. the shoulders around $\sim1$ eV in Fe and Co), thanks to the itinerant electronic bands penetrating the Fermi level.

In order to quantify the effective exchange splitting and Stoner pair gap, we first identify the global maximum of the Stoner spectra in Fig. \ref{fig:stoner_spectrum} along with the corresponding lower and upper half-maxima $\omega_1$ and $\omega_2$. $\Delta_\mathrm{KS}$ and $\Delta_\mathrm{MB}$ are then calculated as the first moment of the respective Stoner spectra in between the two half-maxima,
\begin{equation}\label{eq:stoner pair gap}
    \Delta_\mathrm{MB} = \int_{\omega_1}^{\omega_2}\hbar\omega\:d(\hbar\omega)\:\mathrm{Tr[A^{+-}_\mathrm{S}(\mathbf{q}_\Gamma,\omega)]},
\end{equation}
and listed in Tab. \ref{tab:stoner shift}.
\begin{table}[tb]
    \centering
    \begin{tabular}{c|c|c|c|c} 
        [eV] &
        Fe (bcc) & Ni (fcc) & Co (fcc) & Co (hcp) \\ 
        \hline
        $\Delta_\mathrm{KS}$
        & 2.204 
        & 0.715 
        & 1.714 
        & 1.693 
        \\
        $\Delta_\mathrm{MB}$
        & 1.830 
        & 0.634 
        & 1.530 
        & 1.502 
        \\
        \hline
        shift
        & 0.374 
        & 0.081 
        & 0.184 
        & 0.191 
    \end{tabular}
    \caption{Effective exchange splitting ($\Delta_\mathrm{KS}$) and Stoner pair gap ($\Delta_\mathrm{MB}$) in eV calculated from Eq. \eqref{eq:stoner pair gap}. The shift, $\Delta_\mathrm{KS}-\Delta_\mathrm{MB}$, quantifies the effective electron-hole bonding of the Stoner pairs.}
    \label{tab:stoner shift}
\end{table}
Despite that the Stoner pairs are bonded by exchange (the Dyson Eq. \eqref{eq:ALDA dyson equation} does not include a Hartree/Coulomb term), the Stoner shift (bonding energy) in Fe, Ni and Co is rather significant, ranging from 11\% of the exchange splitting in Ni and Co to 17\% and almost 400 meV in Fe. This redshift appears reminiscent of excitonic effects in insulators, which lowers the optical gap due to the binding energy of excited (spin-neutral) electron-hole pairs, ranging from 0.1 eV in semiconductors to several eV in wide gap insulators \cite{onida_electronic_2002}. In fact, from a many-body point of view both magnons and excitons can be regarded as a coherent superposition of electron hole pairs \cite{Muller2016, olsen_unified_2021} and the binding energy largely originates from the associated localization in real space. However, these two types of excitations have fundamentally different physical origin. While the magnon always exists as a distinct collective quasi-particle in the long wavelength limit, excitons only emerge as distinct quasi-particles (isolated peaks in the spectral function) in cases where the electronic screening is sufficiently low (wide gap insulators or low-dimensional materials). This difference ultimately originates from the fact that the acoustic magnons comprise Goldstone bosons, which is not the case for excitons. Moreover, the binding mechanism is fundamentally different since the magnon binding energy (and redshift of the Stoner continuum) is purely exchange-mediated whereas excitons are bound by direct Coulomb interactions. Finally, we note that TDDFT in the ALDA cannot account for excitonic effects \cite{onida_electronic_2002}, but describes magnon quasi-particles very well.

\subsection{Magnon spectra}

The remainder of the main body of text is dedicated to a detailed analysis of the collective magnon mode lineshapes $a^{+-}_n(\mathbf{q},\omega)$. In the idealized scenario, where each collective mode does not couple to the remaining (collective and single-particle) degrees of freedom---as discussed in Sec. \ref{sec:mode decomposition}---one can effectively regard the Dyson Eq. \eqref{eq:ALDA dyson equation} as a scalar equation. In other words, one may for each collective mode $|v_n(\mathbf{q})\rangle$ try to solve the Dyson Eq. \eqref{eq:ALDA dyson equation} with $v_{n\mathbf{q}}(\mathbf{r})$ as the only basis function,
\begin{equation}
    \chi^{+-}_n(\mathbf{q},\omega) = \frac{\chi^{+-}_{\mathrm{KS},n}(\mathbf{q},\omega)}{1 - \xi^{++}_n(\mathbf{q},\omega)},
\end{equation}
where $\xi^{++}_n(\mathbf{q},\omega)=\langle v_n(\mathbf{q})|\Xi^{++}(\mathbf{q},\omega)|v_n(\mathbf{q})\rangle$. This effectively reduces the mathematical complexity to that of the homogeneous electron gas (HEG). In the HEG, $\Xi^{++}(q,\omega)$ and $\chi^{+-}_\mathrm{KS}(q,\omega)$ differ only by a factor of $f_\mathrm{LDA}^{-+}$ (at the ALDA level), why $\mathrm{Im}\,\chi(q,\omega)=\mathrm{Im}\,\chi_\mathrm{KS}(q,\omega)/|1-\Xi^{++}(q,\omega)|^2$. In order to assess the actual coupling between each of the collective magnon modes and the remaining degrees of freedom, one may therefore compute a singular basis function lineshape after the HEG fashion,
\begin{equation}\label{eq:singular basis function enhancement}
    a^{+-}_{\mathrm{sbf},n}(\mathbf{q},\omega) = \frac{a^{+-}_{\mathrm{KS},n}(\mathbf{q},\omega)}{|1-\xi^{++}_n(\mathbf{q},\omega)|^2},
\end{equation}
and study its deviations from the actual many-body magnon lineshape $a^{+-}_n(\mathbf{q},\omega)$.

\subsubsection{Co (fcc)}\label{sec:fcc-Co results}

\begin{figure*}[tb]
    \centering
    \includegraphics[scale=1.0]{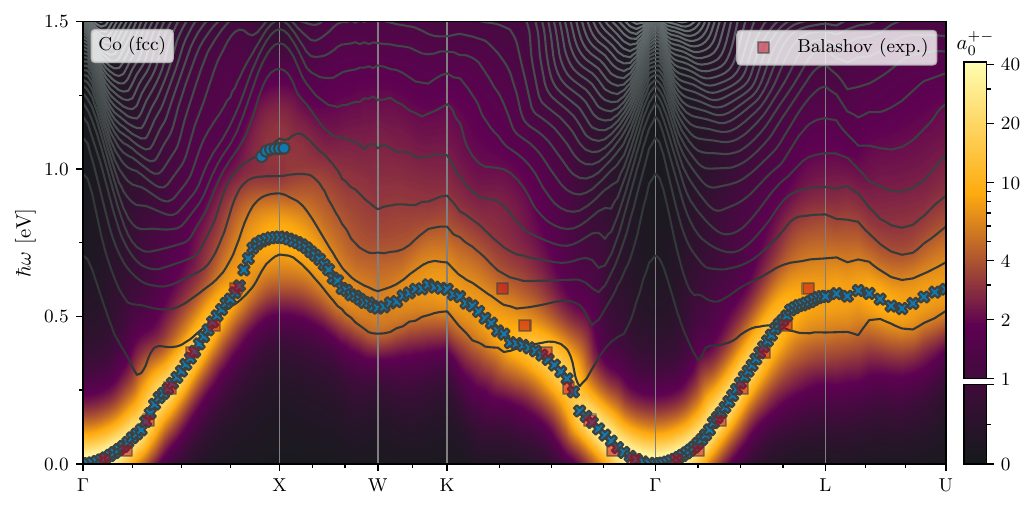}
    \caption{Magnon mode lineshape $a^{+-}_0(\mathbf{q},\omega)$ in fcc-Co (colored contour) overlaid by the corresponding Kohn-Sham lineshape $a^{+-}_{\mathrm{KS},0}(\mathbf{q},\omega)$ (greyscale line contour). Both contours are linear up to $a^{+-}_0=1$ (atomic units), and logarithmic above. Coherent magnon peaks in the many-body lineshape are indicated with crosses, incoherent peaks with disks. The extracted peak positions are compared to inelastic scanning tunnel spectroscopy data (translucent squares) measured on a nine monolayer Co/Cu(100) film assuming isotropy of the dispersion \cite{Balashov2009}.}
    \label{fig:fcc-Co_magnon_spectrum}
\end{figure*}
In Fig. \ref{fig:fcc-Co_magnon_spectrum}, we present the ALDA magnon spectrum of fcc-Co. Across the entire BZ, the magnon mode lineshape is dominated by a main peak feature, dispersing quadratically ($\hbar\omega=D q^2$) near the $\Gamma$-point. As the magnon resonance enters the Stoner continuum, it becomes sensitive to its spectral details, thus inducing anisotropy in the peak dispersion and overall lineshape. Where the peak dispersion is isotropic, agreement with experiment is excellent, but near the BZ boundary the isotropy assumed in the experimental analysis \cite{Balashov2009} clearly breaks down. Furthermore, the ALDA spectrum also exhibits an additional high frequency spectral feature near the X-point, which is likely to be missed when making an isotropic (angular) average.

\begin{figure}
    \centering
    \includegraphics[scale=1.0]{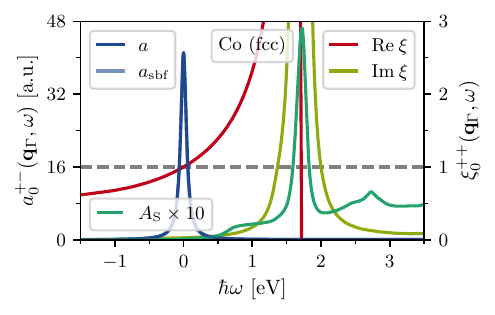}
    \caption{Collective magnon enhancement at $\mathbf{q}_\Gamma$ in Co (fcc). On the left axis, the magnon lineshape $a^{+-}_0(\mathbf{q},\omega)$ (opaque blue) is compared to the singular basis function lineshape \eqref{eq:singular basis function enhancement} (translucent blue) and the many-body Stoner spectrum [trace of Eq. \eqref{eq:stoner continuum spectral function}] (teal). For the $\Gamma$-point in particular, $a^{+-}_{\mathrm{sbf}}$ overlaps perfectly with $a^{+-}$. On the right axis, the magnon mode projection of the self-enhancement function \eqref{eq:magnon self-enhancement function} is shown.}
    \label{fig:fcc-Co_magnon_enhancement}
\end{figure}
In order to understand the rich contents of the full ALDA magnon lineshape in detail (including the additional spectral feature at $\mathbf{q}_\mathrm{X}$), we break down the collective enhancement of the Dyson Eq. \eqref{eq:ALDA dyson equation} and its singular basis function analogue in Eq. \eqref{eq:singular basis function enhancement}, starting with the $\Gamma$-point in Fig. \ref{fig:fcc-Co_magnon_enhancement}. 
At $\mathbf{q}_\Gamma$, $a^{+-}_{\mathrm{sbf},0}$ is practically indistinguishable from the many-body lineshape $a^{+-}_0$. This is---as discussed above---a manifestation of the collective magnon mode decoupling from the remaining single-particle degrees of freedom to an extent where the many-body correlations of the collective mode can be understood based solely on the scalar equation \eqref{eq:singular basis function enhancement}. 
Recalling that the self-enhancement function shares its frequency poles with the Kohn-Sham susceptibility, $\mathrm{Im}\:\xi^{++}_0(\mathbf{q}_\Gamma,\omega)$ is seen to be dominated by a single peak at $\Delta_\mathrm{KS}$. Thanks to the Kramers-Kronig relation, this implies that $\mathrm{Re}\:\xi^{++}_0(\mathbf{q}_\Gamma,\omega)$ changes sign at $\hbar\omega\simeq\Delta_\mathrm{KS}$ and falls off as $1/(\omega-\Delta_\mathrm{KS})$ far from the effective exchange splitting. In particular, $\mathrm{Re}\:\xi^{++}_0(\mathbf{q}_\Gamma,\omega)$ reaches a value of unity at $\omega=0$, thus signifying the existence of a self-sustained collective resonance, appearing as a lorentzian peak in the magnon lineshape.  
Near the Kohn-Sham Stoner peak itself, $|1-\xi^{++}_0(\mathbf{q}_\Gamma,\omega)|\gg1$, meaning that the contribution from $a^{+-}_{\mathrm{KS},0}(\mathbf{q}_\Gamma,\omega)$ effectively is suppressed in Eq. \eqref{eq:singular basis function enhancement}. The self-enhancement function thus provides collective quenching as well as collective enhancement, resulting in a transfer of practically all spectral weight from the Kohn-Sham Stoner resonance 
to the magnon peak. In order to visually separate regions of collective enhancement ($|1-\xi^{++}_n(\mathbf{q},\omega)|<1$) from regions of collective quenching ($|1-\xi^{++}_n(\mathbf{q},\omega)|>1$), it is worth noticing that either $\mathrm{Im}\:\xi^{++}_n(\mathbf{q},\omega)>1$, $\mathrm{Re}\:\xi^{++}_n(\mathbf{q},\omega)<0$ or $\mathrm{Re}\:\xi^{++}_n(\mathbf{q},\omega)>2$ is sufficient to guarantee that $a^{+-}_{\mathrm{sbf},n}(\mathbf{q},\omega)<a^{+-}_{\mathrm{KS},n}(\mathbf{q},\omega)$. 
\begin{figure*}[tb]
    \centering
    \includegraphics[scale=1.0]{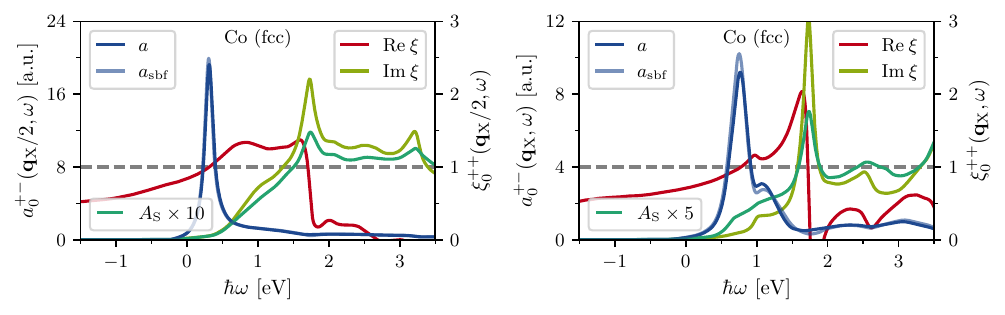}
    \caption{Collective magnon enhancement at $\mathbf{q}_\mathrm{X}/2$ (left) and $\mathbf{q}_\mathrm{X}$ (right) in Co (fcc). 
    }
    \label{fig:fcc-Co_magnon_damping-branching}
\end{figure*}
In essence, this also explains why the bonding nature of the exchange interaction generally results in a redshift of spectral weight through the Dyson Eq. \eqref{eq:ALDA dyson equation}. Namely, it is due to the negative sign of $f_\mathrm{LDA}^{-+}$ that each Kohn-Sham band transition supply a positively/negatively valued contribution to the reactive part of $\Xi^{++}(\mathbf{q},\omega)$ (and therefore $\mathrm{Re}\:\xi^{++}_n(\mathbf{q},\omega)$) below/above the excitation frequency, thus favoring enhancement below and quenching above each spin-flip.

At finite crystal momentum transfers $\hbar\mathbf{q}$, the $\mathrm{Im}\:\xi^{++}_n(\mathbf{q},\omega)$ lineshape broadens, eventually overlapping with the magnon resonance(s) in metals. 
However, this does not necessarily imply a complete breakdown of the quasi-particle picture described in Secs. \ref{sec:lrtddft theory}, \ref{sec:bloch susceptibility theory} and \ref{sec:theory ALDA}. 
An eigenvalue of $1+i\delta$ ($\delta\ll1)$ still signifies the existence of a fluctuational eigenmode which in practice is self-sustained by the many-body electronic correlations, however only for a finite period of time. Peaks in $a^{+-}_n(\mathbf{q},\omega)$ that appear as $\mathrm{Re}\:\xi^{++}_n(\mathbf{q},\omega)$ reaches unity can therefore still be interpreted as coherent magnon resonances despite $\mathrm{Im}\:\xi^{++}_n(\mathbf{q},\omega)>0$, but in this case with finite lifetimes, see also Ref. \cite{Skovhus2024}. 
One such example is found in Fig. \ref{fig:fcc-Co_magnon_damping-branching}, where the magnon enhancement at $\mathbf{q}_\mathrm{X}/2$ in fcc-Co is shown (left panel). Here, $\mathrm{Re}\:\xi^{++}_0(\mathbf{q},\omega)$ crosses unity where $\mathrm{Im}\:\xi^{++}_0(\mathbf{q},\omega)$ is small, but finite, resulting in a broadening of the lineshape (Landau damping), especially towards higher frequencies. The actual magnon mode lineshape is still qualitatively identical to its singular basis function analogue \eqref{eq:singular basis function enhancement}, and the coupling to single-particle degrees of freedom only result in a blueshift of 8 meV in the magnon peak position from $a^{+-}_{\mathrm{sbf},0}(\mathbf{q}_\mathrm{X}/2,\omega)$ to $a^{+-}_0(\mathbf{q}_\mathrm{X}/2,\omega)$. For frequencies above the main magnon peak, $a^{+-}_0(\mathbf{q}_\mathrm{X}/2,\omega)$ diminishes as $\mathrm{Im}\:\xi^{++}_0(\mathbf{q}_\mathrm{X}/2,\omega)$ increases, effectively quenching the scattering amplitude of the collective mode inside the Stoner continuum. It is worth noting that whereas this upper high frequency tail in $a^{+-}_0(\mathbf{q}_\mathrm{X}/2,\omega)$ is physical, the lower lorentzian-like linewidth is instead due mainly to the artificial broadening of $\eta=50$ meV applied to converge the $k$-point integrals of Eqs. \eqref{eq:plane-wave Kohn-Sham susceptibility} and \eqref{eq:plane-wave self-enhancement function}. Because of the inverse relation between Kohn-Sham Stoner weight and lineshape amplitude---namely that factors of $a^{+-}_{\mathrm{KS},n}(\mathbf{q},\omega)/[\mathrm{Im}\:\xi^{++}_n(\mathbf{q},\omega)]^2$ in Eq. \eqref{eq:singular basis function enhancement} scale as $1/a^{+-}_{\mathrm{KS},n}(\mathbf{q},\omega)$---one can generally inspect whether the width of the lineshape is physical or not by checking whether the many-body contours of $a^{+-}_n(\mathbf{q},\omega)$ follow (inversely) those of $a^{+-}_{\mathrm{KS},n}(\mathbf{q},\omega)$. For instance, the linewidths around the $\Gamma$-point in Fig. \ref{fig:fcc-Co_magnon_spectrum} are clearly artificial, whereas lineshapes (especially the upper part) of $q$-points near the BZ boundary are physical.

Finally, we present in Fig. \ref{fig:fcc-Co_magnon_damping-branching} also the magnon enhancement at the X-point (right panel). Here, the main peak in the magnon mode lineshape is accompanied by a secondary peak at a frequency just above. Despite such intricate details, $a^{+-}_{\mathrm{sbf},0}(\mathbf{q}_\mathrm{X},\omega)$ is still qualitatively identical to $a^{+-}_0(\mathbf{q}_\mathrm{X},\omega)$, meaning that the physical nature of e.g. the double-peak structure can be straighforwardly assessed based on the scalar Eq. \eqref{eq:singular basis function enhancement}. In particular, the secondary peak exists due to a shoulder in $\mathrm{Im}\:\xi^{++}_0(\mathbf{q}_\mathrm{X},\omega)$ at around 1 eV. The shoulder is there due to an underlying Stoner pair excitation, providing $\mathrm{Re}\:\xi^{++}_0(\mathbf{q}_\mathrm{X},\omega)$ with a local maximum and minimum below/above its Kohn-Sham excitation frequency. Since $\mathrm{Re}\:\xi^{++}_0(\mathbf{q}_\mathrm{X},\omega)\gtrsim1$ just above the main magnon resonance, this new minimum in $\mathrm{Re}\:\xi^{++}_0(\mathbf{q}_\mathrm{X},\omega)$ imposes a new maximum in the enhancement factor $1/|1-\xi^{++}_0(\mathbf{q}_\mathrm{X},\omega)|^2$, ultimately resulting in a secondary peak of the lineshape. Now, since $\mathrm{Re}\:\xi^{++}_0(\mathbf{q}_\mathrm{X},\omega)$ does not cross unity for the secondary peak, it cannot be interpreted as a coherent magnon peak. However, it is not a Stoner excitation either. In fact, the Stoner excitation responsible for the peak is \textit{redshifted} by the Dyson Eq. \eqref{eq:ALDA dyson equation} and visible as a shoulder just \textit{below} the Kohn-Sham excitation frequency in $A_\mathrm{S}^{+-}(\mathbf{q}_\mathrm{X},\omega)$. The peak in $a^{+-}_0(\mathbf{q}_\mathrm{X},\omega)$ thus corresponds to a \textit{separate} excitation of the many-body system, which neither is a coherent magnon excitation nor a Stoner pair excitation. Since it is a feature of the collective magnon lineshape and exists due to a maximum in the collective enhancement factor, we will refer to it (and other peaks like it) as an incoherent magnon excitation, see also Ref. \cite{Skovhus2024}. Contrary to the coherent magnon excitation, which is blueshifted by 16 meV from the coupling to the single-particle degrees of freedom, the incoherent magnon excitation (peaked at $\hbar\omega=1.069$ eV in $a^{+-}_0(\mathbf{q}_\mathrm{X},\omega)$) is instead redshifted by 27 meV. 
\begin{figure*}[tb]
    \centering
    \includegraphics[scale=1.0]{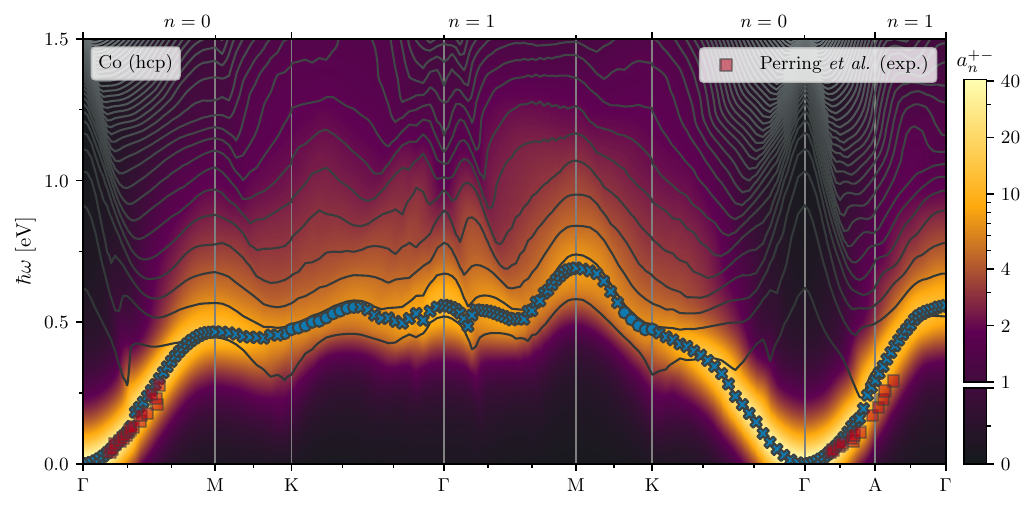}
    \caption{Magnon mode lineshape $a^{+-}_n(\mathbf{q},\omega)$ in hcp-Co (colored contour) overlaid by the corresponding Kohn-Sham lineshape $a^{+-}_{\mathrm{KS},n}(\mathbf{q},\omega)$ (greyscale line contour). Both contours are linear up to $a^{+-}_0=1$ (atomic units), and logarithmic above. The top axis indicates the magnon mode, changing between $n=0$ (acoustic) and $n=1$ (optical) at each K and A point. Coherent magnon peaks in the many-body lineshape are indicated with crosses, incoherent peaks with disks. The extracted peak positions are compared to inelastic neutron scattering data (translucent squares) \cite{Perring1995}.}
    \label{fig:hcp-Co_magnon_spectrum}
\end{figure*}
In addition to the two magnon branches already discussed, the collective magnon mode lineshape at the X-point in Fig. \ref{fig:fcc-Co_magnon_damping-branching} also includes two additional peaks at frequencies above the exchange splitting. Here, $\mathrm{Im}\:\xi^{++}_0(\mathbf{q}_\mathrm{X},\omega)$ is momentarily decreasing, reaching two subsequent valley floors where $\mathrm{Im}\:\xi^{++}_0(\mathbf{q}_\mathrm{X},\omega)<1$ and $0<\mathrm{Re}\:\xi^{++}_0(\mathbf{q}_\mathrm{X},\omega)<2$, thus allowing for the collective enhancement of two incoherent magnon peaks due to corresponding maxima in the enhancement factor $1/|1-\xi^{++}_0(\mathbf{q}_\mathrm{X},\omega)|^2$. Adopting the terminology of Ref. \cite{Skovhus2024}, we will refer to such excitations as valley magnons. By definition, collectively enhanced peaks can only exist above the exchange splitting peak if $\mathrm{Im}\:\xi^{++}_n(\mathbf{q},\omega)$ exhibits at least one secondary peak at an even higher frequency, hence the valley terminology. If not, $\mathrm{Re}\:\xi^{++}_n(\mathbf{q},\omega)<0$ for $\hbar\omega\gtrsim\Delta_\mathrm{KS}$, leading to collective quenching of the high-frequency spectral weight, see e.g. Fig. \ref{fig:fcc-Co_magnon_enhancement}.

\subsubsection{Co (hcp)}\label{sec:hcp-co results}

\begin{figure}
    \centering
    \includegraphics[scale=1.0]{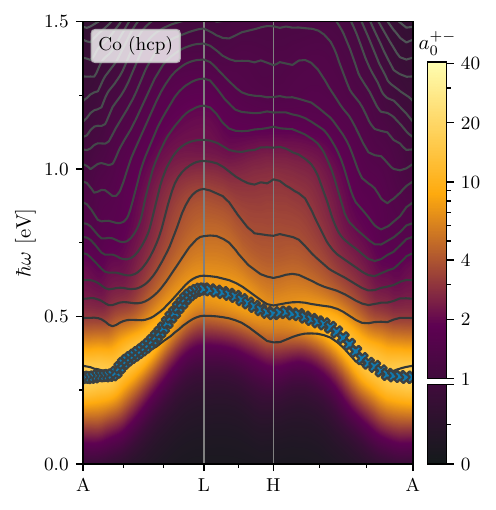}
    \caption{Magnon mode lineshape $a^{+-}_0(\mathbf{q},\omega)$ in hcp-Co at the degenerate hexagonal face of the BZ boundary (colored contour) overlaid by the corresponding Kohn-Sham lineshape $a^{+-}_{\mathrm{KS},0}(\mathbf{q},\omega)$ (greyscale line contour) and extracted coherent magnon peak frequencies (blue crosses). For additional details, please refer to Fig. \ref{fig:hcp-Co_magnon_spectrum}.}
    \label{fig:hcp-Co_degenerate_magnon_spectrum}
\end{figure}
\begin{figure*}[tb]
    \centering
    \includegraphics[scale=1.0]{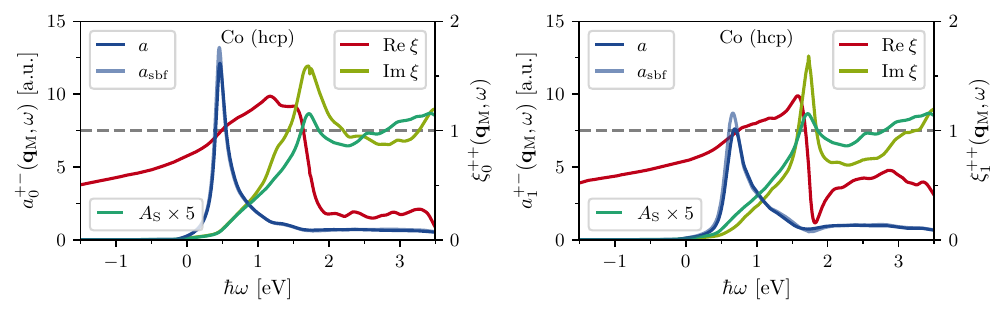}
    \caption{Collective magnon enhancement of the acoustic mode (left) and the optical mode (right) at $\mathbf{q}_\mathrm{M}$ in Co (hcp). 
    }
    \label{fig:hcp-Co_M0-M1_magnon_coherency}
\end{figure*}
In Figs. \ref{fig:hcp-Co_magnon_spectrum} and \ref{fig:hcp-Co_degenerate_magnon_spectrum}, we present the magnon spectrum of hcp-Co. Although the majority spectral function involves two collective modes (an acoustic and an optical mode), the total magnon spectrum remains gapless as a whole, namely because the two modes are degenerate along the edges of the BZ boundary (including the K-point) and on its hexagonal face (including the A-point) \cite{Scheie2022}. The symmetrical equivalence of the two modes at the degenerate high-symmertry points allows us to depict their lineshapes as a single contour map in Fig. \ref{fig:hcp-Co_magnon_spectrum}, with nondegenerate high-symmetry points $\Gamma$ and M appearing twice, once for each of the modes. Similarly to fcc-Co, the acoustic magnon dispersion agrees well with experiment for the undamped (quadratic) part of the dispersion. However, for both the $\Gamma$-M and $\Gamma$-A directions, deviations start occurring once the magnon resonance enters the Stoner continuum. This is almost inevitable. In the direct vicinity of a magnon resonance, tiny features in the Stoner continuum can greatly impact the magnon mode lineshape, ultimately leading to secondary magnon branches, such at for the X-point in fcc-Co (Fig. \ref{fig:fcc-Co_magnon_damping-branching}). In order to match experiment exactly, it is therefore not sufficient to accurately capture the coarse grained dispersion of the point where $\mathrm{Re}\:\xi^{++}_n(\mathbf{q},\omega)=1$. One would also need to accurately resolve all Kohn-Sham band transitions in its near vicinity. This is a hard ask for any \textit{ab initio} theory, also beyond the LDA level.

Among all the high-symmetry points on the BZ boundary, it is only at the M-point where the two modes are nondegenerate. That being said, the magnon mode lineshapes are still qualitatively rather similar, as illustrated in Fig. \ref{fig:hcp-Co_M0-M1_magnon_coherency}. Comparing $\mathrm{Im}\:\xi^{++}_0(\mathbf{q}_\mathrm{M},\omega)$ to $\mathrm{Im}\:\xi^{++}_1(\mathbf{q}_\mathrm{M},\omega)$, it is clear that both are composed out of the same Kohn-Sham Stoner pairs, but with differences in the spectral weight of each excitation. The difference in weights ultimately results in different magnon peak positions, that is, $\mathrm{Re}\:\xi^{++}_0(\mathbf{q}_\mathrm{M},\omega)$ and $\mathrm{Re}\:\xi^{++}_1(\mathbf{q}_\mathrm{M},\omega)$ cross unity at different frequencies. Once again, the singular basis function lineshapes \eqref{eq:singular basis function enhancement} are qualitatively identical to the actual magnon lineshapes. However, whereas the magnon peak position of the acoustic mode is blueshifted by only 9 meV from the coupling to the single-particle degrees of freedom, the optical mode is shifted by 27 meV. This difference simply reflects that the optical magnon resides at a higher energy, where the amplitude of the many-body Stoner spectrum is larger.

\begin{figure}
    \centering
    \includegraphics[scale=1.0]{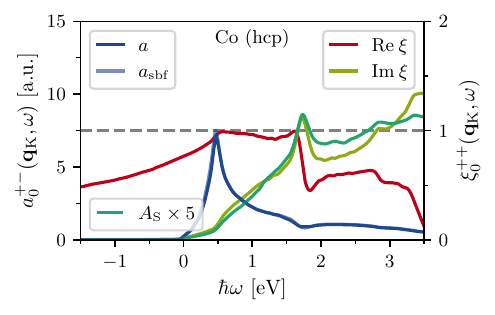}
    \caption{Collective magnon enhancement at $\mathbf{q}_\mathrm{K}$ in Co (hcp). 
    }
    \label{fig:hcp-Co_K_magnon_incoherency}
\end{figure}
Finally, we note that some of the magnon peak positions of Fig. \ref{fig:hcp-Co_magnon_spectrum} have been marked as incoherent, meaning that $\mathrm{Re}\:\xi^{++}_n(\mathbf{q},\omega)$ reaches a maximum near unity without actually crossing it. Most prominently, this is the case for the K-point, whose magnon enhancement is illustrated in Fig. \ref{fig:hcp-Co_K_magnon_incoherency}. Here, the spectral weight in $\mathrm{Im}\:\xi^{++}_0(\mathbf{q},\omega)$ has been broadened so much that $\mathrm{Re}\:\xi^{++}_0(\mathbf{q},\omega)$ never crosses unity, but approaches it instead very closely. The magnon peak can therefore be characterized as near coherent, and it is quite likely that an elimination of the $\eta=50$ meV artificial broadening (e.g. by analytic continuation) would lead to a recharacterization of the peak as coherent. In addition to the near-crossing at the magnon peak, one may notice that $\mathrm{Re}\:\xi^{++}_0(\mathbf{q}_\mathrm{K},\omega)$ also approaches unity close to the exchange splitting peak itself. However, whereas the actual magnon peak resides at a frequency where the Stoner spectrum is low of amplitude (leading to a blueshift of only 6 meV from the coupling to single-particle degrees of freedom), the upper approach towards $\mathrm{Re}\:\xi^{++}_0(\mathbf{q}_\mathrm{K},\omega)\simeq1$ happens in the near vicinity of where $\mathrm{Im}\:\xi^{++}_0(\mathbf{q}_\mathrm{K},\omega)$ crosses unity, thus favoring collective quenching over collective enhancement.

\subsubsection{Fe (bcc)}

\begin{figure*}[tb]
    \centering
    \includegraphics[scale=1.0]{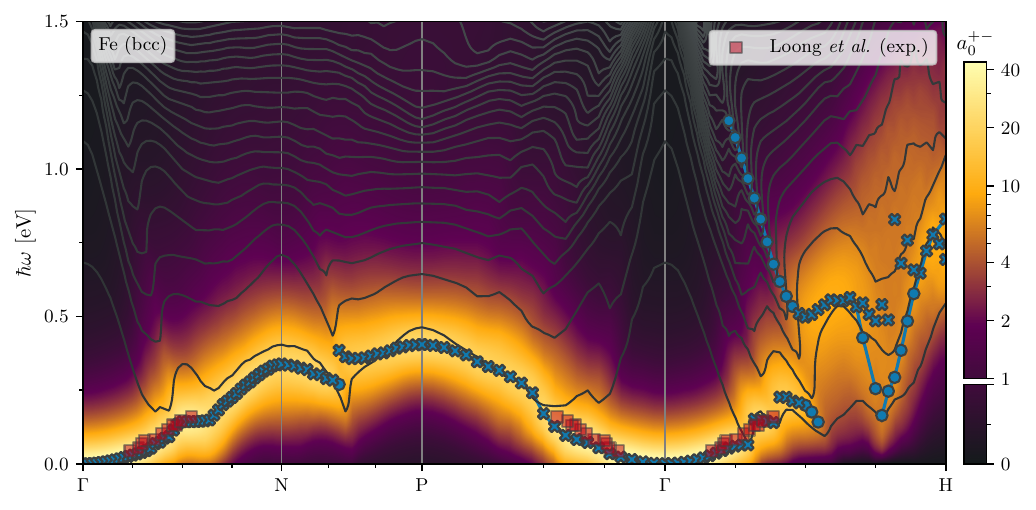}
    \caption{Magnon mode lineshape $a^{+-}_0(\mathbf{q},\omega)$ in bcc-Fe (colored contour) overlaid by the corresponding Kohn-Sham lineshape $a^{+-}_{\mathrm{KS},0}(\mathbf{q},\omega)$ (greyscale line contour). Both contours are linear up to $a^{+-}_0=1$ (atomic units), and logarithmic above. Coherent magnon peaks in the many-body lineshape are indicated with crosses, incoherent peaks with disks. The extracted peak positions are compared to inelastic neutron scattering data (translucent squares) assuming isotropy of the dispersion \cite{Loong1984}.}
    \label{fig:bcc-Fe_magnon_spectrum}
\end{figure*}
In Fig. \ref{fig:bcc-Fe_magnon_spectrum}, we present the magnon spectrum in bcc-Fe. Along the $\Gamma$-N and $\Gamma$-P directions, the magnon dispersion is reasonably simple. However, on the N-P path and along the $\Gamma$-H direction one observes a series of apparent discontinuities, while the lineshape near the H-point is very broad with a plethora of local peaks. In literature, the broadening near the H-point is sometimes interpreted as the disappearance of the collective magnon \cite{Buczek2011b,Cao2017,Skovhus2021} resonance, leaving only Stoner pair excitations. Contrary to this interpretation, we observe a clear separation of the collective magnon mode from the remaining single-particle Stoner modes in the eigendecomposition of the majority spectral function, also at the H-point. Moreover, $\mathrm{Re}\:\xi^{++}_0(\mathbf{q},\omega)$ crosses unity at least once for all the recorded wave vectors $\mathbf{q}$. In Fig. \ref{fig:bcc-Fe_magnon_spectrum}, we show all the corresponding coherent magnon peaks, but leave some of the incoherent peaks unannotated in order not to obscure the overall trends.
\begin{figure}
    \centering
    \includegraphics[scale=1.0]{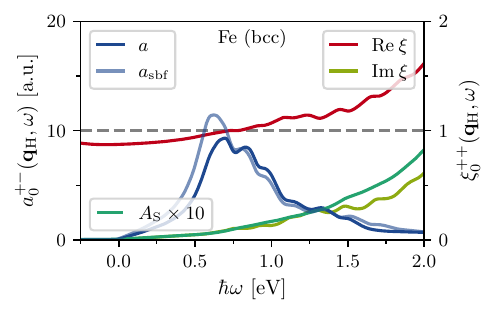}
    \caption{Collective magnon enhancement at $\mathbf{q}_\mathrm{H}$ in Fe (bcc). 
    }
    \label{fig:bcc-Fe_magnon_coherency}
\end{figure}
\begin{figure*}[tb]
    \centering
    \includegraphics[scale=1.0]{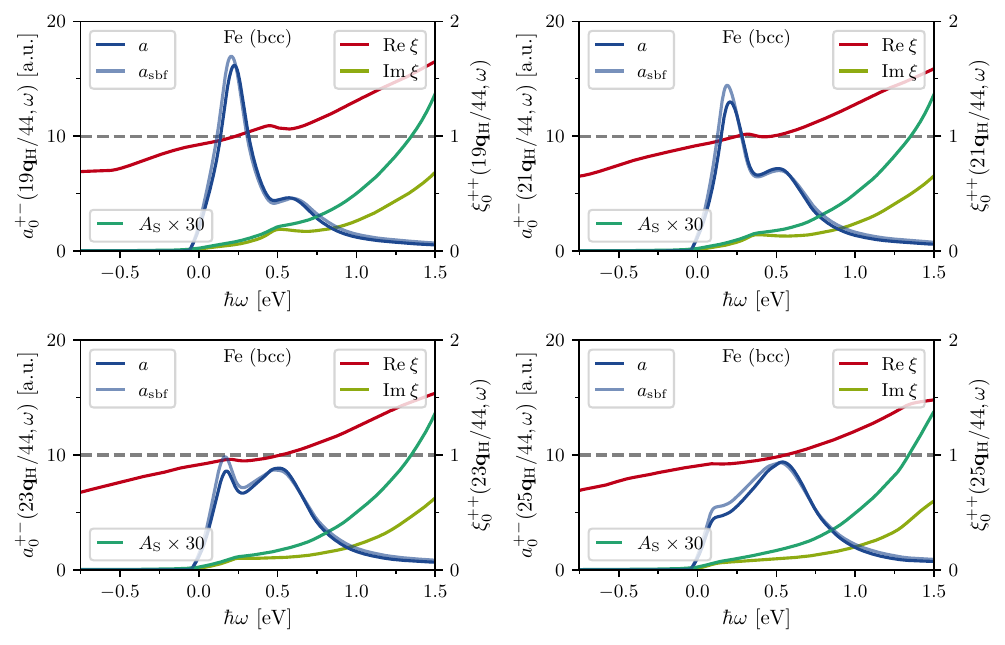}
    \caption{Collective magnon enhancement around $\mathbf{q}_\mathrm{H}/2$ in Fe (bcc). 
    }
    \label{fig:bcc-Fe_magnon_all-branching}
\end{figure*}
Exemplified by the magnon enhancement at the H-point in Fig. \ref{fig:bcc-Fe_magnon_coherency}---where the singular basis function lineshape \eqref{eq:singular basis function enhancement} once again explains the existence of each individual peak---our interpretation is instead as follows. As the overall magnon resonance (point where $\mathrm{Re}\:\xi^{++}_0(\mathbf{q},\omega)=1$) enters the Stoner continuum, the lineshape is broadened with a long high-frequency tail, in complete analogy to e.g. the spectral broadening at the M-point in hcp-Co shown in Fig. \ref{fig:hcp-Co_M0-M1_magnon_coherency}. However, because individual sets of Kohn-Sham band transitions have made $\mathrm{Im}\:\xi^{++}_0(\mathbf{q}_\mathrm{H},\omega)$ rugged rather than smooth, $\mathrm{Re}\:\xi^{++}_0(\mathbf{q}_\mathrm{H},\omega)$ is rugged as well, meaning that the collective enhancement factor $1/|1-\xi^{++}_0(\mathbf{q}_\mathrm{H},\omega)|^2$ goes through a series of maxima resulting in each their own peak in the lineshape. All of these peaks correspond to distinct excitations of the many-body system with collective magnon character (some coherent, some incoherent). In addition, the Kohn-Sham Stoner features present in $\mathrm{Im}\:\xi^{++}_0(\mathbf{q},\omega)$ also provide spectral weight to the many-body Stoner continuum $A_\mathrm{S}^{+-}(\mathbf{q},\omega)$, but at different frequencies than in the collective magnon mode, corresponding therefore to a different set of many-body excitations with single-particle Stoner pair character. This is, of course, a very complicated picture. However, from a practical point of view, it is important to emphasize that all the magnon excitations at a given $\mathbf{q}$ belong to the same collective mode of the majority spectral function $A^{+-}(\mathbf{q},\omega)$. This means that they invoke the same material response (spatially) to weak external perturbations through Eq. \eqref{eq:linear response relation}, simply modulating the amplitude of the response as a function of frequency. It might therefore seem quite sensible to coarse grain the 
spectral picture and interpret all the individual magnon peaks in Fig. \ref{fig:bcc-Fe_magnon_coherency} as forming a single magnon quasi-particle with a short lifetime according to the overall broadening.

While the single quasi-particle interpretation is practical from a modeling perspective (e.g. in order to map the problem onto a Heisenberg model), it is also comes at the cost of completeness. For instance, one would miss the fact that both the largest and second largest spectral peak at $\mathbf{q}_\mathrm{H}$ exist due to an associated coherent $\mathrm{Re}\:\xi^{++}_0(\mathbf{q},\omega)=1$ crossing, each with a widely different peak renormalization from the coupling to single-particle degrees of freedom (the lower peak is blueshifted by 71 meV, the upper peak by 28 meV).
In fact, the coexistence of coherent magnon peaks means that there is no unique way to define "the" magnon dispersion in bcc-Fe, and it is only when one tries to follow the coherent peaks (crosses) in Fig. \ref{fig:bcc-Fe_magnon_spectrum} that the dispersion appears discontinuous. Instead of extracting a nonunique and discontinuous single quasi-particle dispersion, one may instead try to organize the peaks into separate (sometimes overlapping) magnon branches, each appearing only at a subset of the BZ. Doing so results in branches that appear both continuous and smooth, as illustrated in Fig. \ref{fig:bcc-Fe_magnon_spectrum} with lines connecting the discernible branches.

For bcc-Fe, the most prominent example of magnon branch coexistence appears in the vicinity of $\mathbf{q}_\mathrm{H}/2$, as illustrated in Fig. \ref{fig:bcc-Fe_magnon_all-branching}, see also Refs. \cite{Buczek2011b,Friedrich2014,SkovhusPhD}. At $\mathbf{q}_\mathrm{H}/2$, a Stoner peak with negative dispersion (the Stoner peak frequency decreases with $|\mathbf{q}|$) crosses the magnon resonance, before turning into a minority Stoner excitation at negative frequencies beyond $\mathbf{q}_\mathrm{H}/2$. Analogous to the magnon branching at the X-point in fcc-Co (see Fig. \ref{fig:fcc-Co_magnon_damping-branching}), the Stoner feature supplies $\mathrm{Re}\:\xi^{++}_0(\mathbf{q},\omega)$ with positive weight (favoring enhancement) at frequencies below the Stoner peak and negative weight (favoring quenching) for frequencies above the peak, thus forming a frequency gap between the two magnon branches. In this way, the Stoner feature may be seen to provide the collective magnon subspace (which includes multiple many-body eigenstates) with a coupling term, that splits up the spectral weight into a "bonding" and an "antibonding" magnon branch. 

\begin{figure*}[tb]
    \centering
    \includegraphics[scale=1.0]{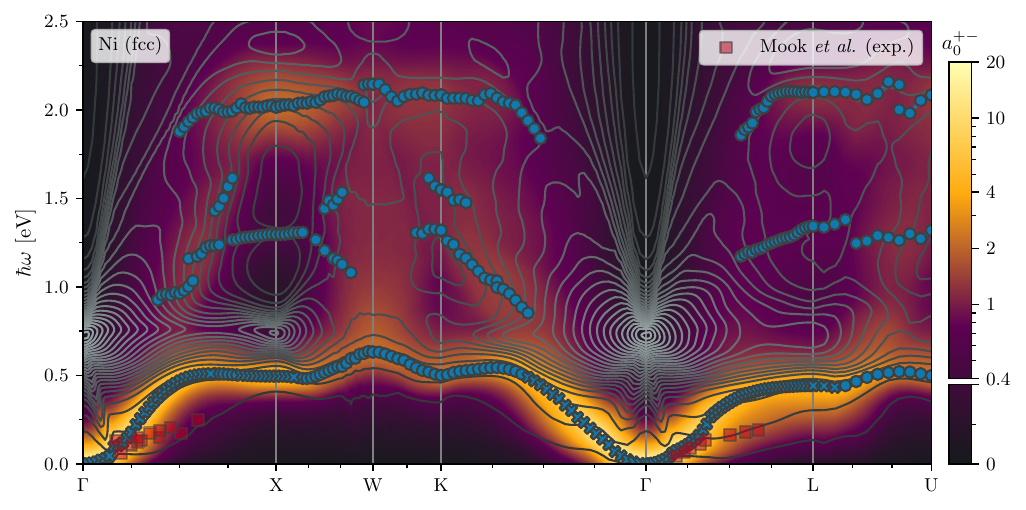}
    \caption{Magnon mode lineshape $a^{+-}_0(\mathbf{q},\omega)$ in fcc-Ni (colored contour) overlaid by the corresponding Kohn-Sham lineshape $a^{+-}_{\mathrm{KS},0}(\mathbf{q},\omega)$ (greyscale line contour). Both contours are linear up to $a^{+-}_0=0.4$ (atomic units), and logarithmic above. Coherent magnon peaks in the many-body lineshape are indicated with crosses, incoherent peaks with disks. The extracted peak positions are compared to inelastic neutron scattering data \cite{Mook1985}.}
    \label{fig:fcc-Ni_magnon_spectrum}
\end{figure*}
The coexistence of multiple magnon branches within a single magnon mode is a hallmark of itinerant ferromagnetism, and it should not come as a surprise that one ends up with a nonanalytic magnon dispersion when neglecting it, e.g. when mapping the problem onto a Heisenberg model using the magnetic force theorem \cite{Durhuus2023}.
In particular, it is noteworthy that the appearance of Stoner peaks with negative dispersion---also called Stoner stripes, see Ref. \cite{Friedrich2020}---is a trait all four materials studied here have in common [look for downwards-pointing bulges in the Kohn-Sham lineshape contours $a^{+-}_{\mathrm{KS},n}(\mathbf{q},\omega)$]. Every time such a Stoner stripe crosses the overall magnon dispersion, one observes traces of magnon branch coexistence. However, due to the artificial spectral broadening of $\eta=50$ meV, we are only able to discern both peaks at the same $\mathbf{q}$ for a few instances, meaning that the Stoner stripes mainly provide the magnon peak dispersion with a nonanalytical visual quality. 
This includes the remaining discontinuities along the N-P and $\Gamma$-H paths in bcc-Fe (see Fig. \ref{fig:bcc-Fe_magnon_spectrum}), but also the Co dispersions presented above. For hcp-Co in Fig. \ref{fig:hcp-Co_magnon_spectrum}, Stoner stripes induce nonanalyticity along the $\Gamma$-M and $\Gamma$-A directions of the acoustic mode and along the $\Gamma$-K and $\Gamma$-M directions of the optical mode, while the dispersion of fcc-Co in Fig. \ref{fig:fcc-Co_magnon_spectrum} exhibits a nonanalytical point along the $\Gamma$-K direction. It should be emphasized (see also Sec. \ref{sec:hcp-co results}), that the exact $(\mathbf{q},\omega)$ coordinates of the Stoner stripe crossings probably are not accurate at the ALDA level. In particular, one should not be discouraged by the comparison to experiment in Fig. \ref{fig:bcc-Fe_magnon_spectrum}, which is actually quite good given that the experimental analysis assumes isotropy of the dispersion.

\subsubsection{Ni (fcc)}\label{sec:ni results}

Finally, we present in Fig. \ref{fig:fcc-Ni_magnon_spectrum} the ALDA magnon spectrum of fcc-Ni. Similar to Fe and Co, Ni exhibits clear traces of magnon branch coexistence due to Stoner stripes---see also Refs. \cite{Karlsson2000,SasIoglu2010,Buczek2011b,Cao2017,Friedrich2020,Skovhus2021}---but in this case very close to the $\Gamma$-point. In fact, magnon branch coexistence has been observed in Ni along the $\Gamma$-X direction also experimentally \cite{Mook1985}. Contrary to Fe and Co, however, ALDA does not match the experimental magnon dispersion very well. This is mainly attributed a lacking description of the coupling between transverse magnetic excitations and the electronic bands at the LDA level, resulting in an overestimation of the exchange splitting \cite{Muller2016,Muller2019,Nabok2021,Paischer2023}. Since both ALDA shortcomings and Stoner stripes have been discussed in detail for Ni in literature, we will here focus on other aspects of the spectrum, namely the magnon decoherence and pronounced valley magnon features observed in Fig. \ref{fig:fcc-Ni_magnon_spectrum}.

At the $\Gamma$-point, all spectral weight of the Kohn-Sham lineshape $a^{+-}_{\mathrm{KS},0}(\mathbf{q}_\Gamma,\omega)$ is concentrated at the exchange splitting $\Delta_\mathrm{KS}$, see also Fig. \ref{fig:FeNi_eigendecomposition}. With growing crystal momentum transfer $\hbar q$, the spectral weight spreads out, in some cases forming new sets of peaks in the Kohn-Sham lineshape above the exchange splitting, visible in Fig. \ref{fig:fcc-Ni_magnon_spectrum} as hills in the Kohn-Sham contour. These upper Stoner peaks are responsible for the many incoherent valley magnon branches observed in Ni.
\begin{figure*}[tb]
    \centering
    \includegraphics[scale=1.0]{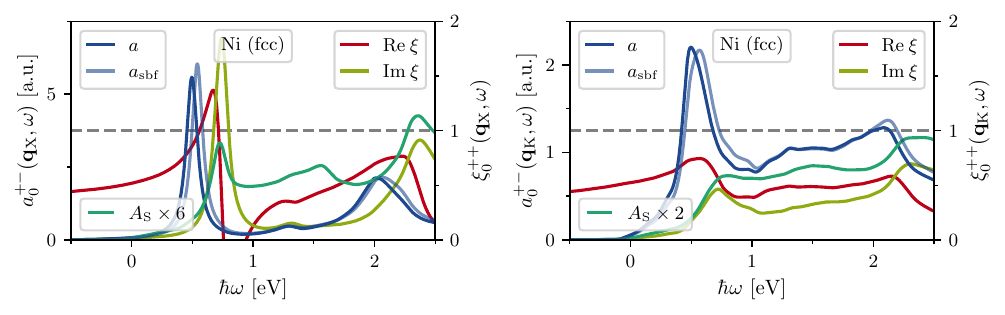}
    \caption{Collective magnon enhancement at $\mathbf{q}_\mathrm{X}$ (left) and $\mathbf{q}_\mathrm{K}$ (right) in Ni (fcc).}
    \label{fig:fcc-Ni_X-K_valley_magnons}
\end{figure*}
The most prominent example of valley magnon enhancement happens at the X-point, as illustrated in the left panel of Fig. \ref{fig:fcc-Ni_X-K_valley_magnons}. Here, $\mathrm{Im}\:\xi^{++}_0(\mathbf{q}_\mathrm{X},\omega)$ has developed a peak at 2.373 eV, giving rise to a new maximum in the enhancement factor $1/|1-\xi^{++}_0(\mathbf{q}_\mathrm{X},\omega)|^2$ in the valley below the peak. Since $\mathrm{Im}\:\xi^{++}_0(\mathbf{q}_\mathrm{X},\omega)\ll1$ in this valley, a clear valley magnon feature forms in the many-body lineshape, peaked at 2.016 eV well below the corresponding many-body Stoner peak at 2.357 eV in $A^{+-}_\mathrm{S}(\mathbf{q}_\mathrm{X},\omega)$. This clear separation of spectral weight underlines that the valley magnon is collective in nature and unambiguously separable from the continuum of Stoner pair excitations.

Despite the spectral broadening of the Kohn-Sham Stoner continuum, $\mathrm{Im}\:\xi^{++}_0(\mathbf{q},\omega)$ still retains a clear exchange splitting peak at the X-point, resulting in a coherent magnon excitation below it. However, this is not the case for many other $q$-points along the BZ boundary, as illustrated by the K-point in the right panel of Fig. \ref{fig:fcc-Ni_X-K_valley_magnons}. Here, the peak at $\Delta_\mathrm{KS}$ in $\mathrm{Im}\:\xi^{++}_0(\mathbf{q}_\mathrm{K},\omega)$ has lost so much spectral weight that $\mathrm{Re}\:\xi^{++}_0(\mathbf{q}_\mathrm{K},\omega)$ never crosses unity, implying that the lower magnon branch is incoherent. Whereas the incoherent $q$-points in hcp-Co (see Sec. \ref{sec:hcp-co results}) probably stem from the artificial broadening applied, the lower magnon branch of fcc-Ni is much less likely to be recharacterized as coherent upon analytic continuation. In fact, the flatness of $\mathrm{Im}\:\xi^{++}_0(\mathbf{q}_\mathrm{K},\omega)$ that results in decoherence of the primary magnon branch also means that an unusually wide frequency window is collectively enhanced rather than quenched, effectively allowing the valley magnons to carry much of the spectral weight. This loss of spectral weight to the valley magnons could have real consequences for derived physical properties that normally are calculated using a single quasi-particle approximation. For instance, one would expect the Curie temperature to be underestimated if all the spectral weight is assigned to the lower magnon branch, e.g. in a Heisenberg model treatment.

Lastly, one may in Fig. \ref{fig:fcc-Ni_X-K_valley_magnons} note that the primary lower magnon branch at both X and K is \textit{redshifted} (by 45 meV and 72 meV respectively) from the coupling to the single-particle degrees of freedom, contrary to the highlighted $q$-points in Fe and Co which are \textit{blueshifted}. In addition, one may also note that Fe and Ni generally seem to undergo shifts of a larger magnitude than in Co, thus explaining the increased basis set sensitivity discussed in Appendix \ref{app:magnon dispersion convergence}. All in all, it seems likely that some of the quantitative discrepancies between (A)LDA magnon dispersions found in literature \cite{Skovhus2021} exist due to insufficient basis set convergence, and, by extension, an insufficient treatment of the coupling to single-particle degrees of freedom. To improve convergence in future implementations, a detailed analysis of the Stoner modes---of which there are multiple with similar amplitudes, see Figs. \ref{fig:Co_eigendecomposition} and \ref{fig:FeNi_eigendecomposition}---is highly warranted. Ideally, one would systematically construct basis functions that closely resemble the most prominent Stoner modes and add them based on the spectral weight of each mode.

\section{Conclusion and Outlook}\label{sec:conclusion}
We have presented a new implementation of LR-TDDFT in the ALDA, which bypasses the numerical inconsistencies that typically lead to a Goldstone gap error. With the magnon branches correctly positioned with respect to the Stoner continuum, we have analyzed the interaction between magnons and noninteracting Stoner pairs in detail for the prototypical ferromagnets bcc-Fe, fcc-Ni, fcc-Co and hcp-Co. In particular, it was shown that the collective character of a given excitation can be quantified by analysis of the associated self-enhancement function. 
A magnon can be characterized as coherent if the real part of the self-enhancement eigenvalue exhibits a unity crossing at the magnon peak, while the Landau damping (and magnon lifetime) is directly determined by the imaginary part. We discussed the concept of incoherent magnons (strongly enhanced, but incoherent peaks in the magnon lineshape), Stoner stripes and valley magnons, which all comprise distinct features in the spectral function that can be directly understood from the self-enhancement function. Finally, the eigenmode analysis allowed us to define a many-body spectral function for the Stoner excitations and thereby quantify the many-body exchange splitting as determined by TDDFT in the ALDA.

The present work paves the way for a detailed understanding of magnetic excitations in itinerant ferromagnets, which exhibit much richer spectral features compared to insulators. While we have focused on the simplest possible ferromagnets, we emphasize that analysis of the self-enhancement function is a general tool that may be used to quantify the collective nature of any spectral feature within the framework of TDDFT.
In addition, most permanent magnets in modern applications are metallic, and understanding these (and predicting new ones) relies on a proper theoretical framework such as the one presented here. In particular, the thermal properties are largely governed by the transverse magnetic susceptibility, and access to this constitutes a minimal requirement for any attempt at predicting Curie temperatures in metals. Previous attempts at predicting critical temperatures in itinerant magnets have mostly been based on Heisenberg model approaches \cite{pajda_ab_2001, bruno_exchange_2003, pavizhakumari_beyond_2025}, which appear rather dubious given the significant deviation of the spectral functions calculated here from isolated magnon branches. While a full temperature-dependent calculation of the transverse magnetic susceptibility currently is out of reach, it is expected that previous estimates of Curie temperatures in itinerant magnets can be significantly improved by including a few dynamical eigenvalues of the full (zero-temperature) spectral function in the thermal modeling. 

\begin{acknowledgments}
The work presented here is supported by the Carlsberg Foundation Grant No. CF24-1413, and the Villum Foundation Grant No.~00029378. Computations were performed on Nilfheim, a high performance computing cluster at the Technical University of Denmark. In this regard, we acknowledge support from the Novo Nordisk Foundation Data Science Research Infrastructure 2022 Grant: A high-performance computing infrastructure for data-driven research on sustainable energy materials, Grant No. NNF22OC0078009.
\end{acknowledgments}

\section*{Data availability}
The data produced for this paper, including the GPAW ground states on which the LR-TDDFT calculations are based, are available upon reasonable request via the authors.

\appendix

\section{Generalized Bloch susceptibility}\label{app:bloch susceptibilities}

\subsection{Generalized susceptibility}

The generalized retarded susceptibility is defined via the Kubo formula \cite{Kubo1957}. Given a set of system operators $\hat{A}$ and $\hat{B}$,
\begin{equation}
    \chi_{BA}(t-t') = -\frac{i}{\hbar} \theta(t-t') 
    \langle [ \hat{B}_0(t), \hat{A}_0(t') ] \rangle_0,
\end{equation}
where $\hat{A}_0(t)=e^{i\hat{H}_0t/\hbar}\hat{A}e^{-i\hat{H}_0t/\hbar}$ and the expectation value $\langle\cdot\rangle_0$ is taken with respect to the unperturbed system (at thermal equilibrium). As a retarded correlation function, $\chi_{BA}$ is analytic in the upper-half of the complex frequency plane and can be written in terms of the system eigenstates $\hat{H}_0 |\alpha\rangle = E_\alpha |\alpha\rangle$,
\begin{equation}
    \chi_{BA}(z) 
    = \sum_{\alpha,\alpha'} \langle \alpha|\hat{B}|\alpha'\rangle\langle \alpha'|\hat{A}|\alpha\rangle \frac{n_\alpha - n_{\alpha'}}{\hbar z - (E_{\alpha'} - E_\alpha)},
    \label{eq:generalized susceptibility lehmann}
\end{equation}
where $n_\alpha$ are the population factors at thermal equilibrium and $z=\omega + i\eta$ with $\eta>0$. The complex frequency poles which constitute the generalized susceptibility \eqref{eq:generalized susceptibility lehmann} can be split into their real and imaginary parts 
by splitting the susceptibility into its reactive and dissipative parts \cite{Jensen1991},
%
\begin{subequations}
   \begin{equation}
       \chi_{BA}(z) = \chi_{BA}'(z) + i \chi_{BA}''(z),
   \end{equation}
   where
   \begin{equation}
       \chi_{BA}'(z) = \frac{1}{2}\left[\chi_{BA}(z) + \chi_{AB}(-z^*)\right]
   \end{equation}
   and
   \begin{equation}
       \chi_{BA}''(z) = \frac{1}{2i}\left[\chi_{BA}(z) - \chi_{AB}(-z^*)\right].
   \end{equation}
\end{subequations}
In this way, the dissipative part of the susceptibility characterizes the eigenstate transitions available to the system at thermal equilibrium in the form of a weighted joint density of states,
\begin{equation}
    \chi''_{BA}(\omega+i\eta) = \sum_{\alpha,\alpha'} \frac{ \langle \alpha|\hat{B}|\alpha'\rangle\langle \alpha'|\hat{A}|\alpha\rangle (-\hbar \eta) (n_\alpha - n_{\alpha'})}{(\hbar \omega - [E_{\alpha'} - E_\alpha])^2 + (\hbar\eta)^2}.
\end{equation}

\subsection{Bloch lattice Fourier transform}

In systems with discrete translational invariance $[\hat{T}_\mathbf{R},\hat{H}_0]=0$, the susceptibility \eqref{eq:generalized susceptibility lehmann} may be cast as a sum over the (energy) eigenstates of lattice translations $\hat{T}_\mathbf{R}|\alpha\rangle=e^{i\mathbf{k}_\alpha\cdot\mathbf{R}}|\alpha\rangle$. Matrix elements involving operators $\hat{A}=\hat{A}(\mathbf{r})$ can then be written in the form of Bloch waves,
\begin{equation}
    \langle \alpha'|\hat{A}(\mathbf{r})|\alpha\rangle = \frac{\Omega_\mathrm{cell}}{\Omega} e^{-i\mathbf{q}_{\alpha'\alpha}\cdot\mathbf{r}} a_{\alpha'\alpha}(\mathbf{r}),
\end{equation}
where $\Omega_\mathrm{cell}$ and $\Omega$ are the unit cell and crystal volumes. The wave vector $\mathbf{q}_{\alpha'\alpha}=-\mathbf{q}_{\alpha\alpha'}$ yields the primitive difference in crystal momentum between states $|\alpha'\rangle$ and $|\alpha\rangle$, while the periodic part of the matrix element $a_{\alpha'\alpha}(\mathbf{r})=a_{\alpha'\alpha}(\mathbf{r}+\mathbf{R})$ is normalized to the unit cell, see also Ref. \cite{Skovhus2021}. Resultantly, the generalized susceptibility \eqref{eq:generalized susceptibility lehmann} becomes periodic on the crystal lattice $\chi_{BA}(\mathbf{r}, \mathbf{r}', z)=\chi_{BA}(\mathbf{r}+\mathbf{R}, \mathbf{r}'+\mathbf{R}, z)$ and subject to the Bloch lattice Fourier transform,
\begin{equation}
    \bar{\chi}_{BA}(\mathbf{r},\mathbf{r}',\mathbf{q},z) = e^{-i\mathbf{q}\cdot(\mathbf{r}-\mathbf{r}')} \sum_{\mathbf{R}'} e^{i\mathbf{q}\cdot\mathbf{R}'} \chi_{BA}(\mathbf{r},\mathbf{r}'+\mathbf{R}',z),
\end{equation}
\begin{equation}
    \chi_{BA}(\mathbf{r},\mathbf{r}',z) = \frac{1}{N_q} \sum_{\mathbf{q}} e^{i\mathbf{q}\cdot(\mathbf{r}-\mathbf{r}')} \bar{\chi}_{BA}(\mathbf{r},\mathbf{r}',\mathbf{q},z).
\end{equation}
The resulting generalized Bloch susceptibility $\bar{\chi}_{BA}$ is periodic in both $\mathbf{r}$ and $\mathbf{r}'$ entries independently, and can be neatly cast in terms of the periodic parts of the matrix elements \cite{SkovhusPhD},
\begin{align}
    \bar{\chi}_{BA}(\mathbf{r},\mathbf{r}',\mathbf{q},z) 
    = \frac{\Omega_\mathrm{cell}}{\Omega} \sum_{\alpha,\alpha'} & b_{\alpha\alpha'}(\mathbf{r}) a_{\alpha'\alpha}(\mathbf{r}') \delta_{\mathbf{q},\mathbf{q}_{\alpha'\alpha}}
    \nonumber \\
    &\times 
    \frac{n_\alpha - n_{\alpha'}}{\hbar z - (E_{\alpha'} - E_\alpha)}.
    \label{eq:generalized bloch susceptibility lehmann}
\end{align}

\subsection{Matrix representation}

Given some orthonormal basis for the periodic degrees of freedom,
\begin{equation}
    f_i(\mathbf{r}) = f_i(\mathbf{r}+\mathbf{R}), \quad \int_{\Omega_\mathrm{cell}} d\mathbf{r}\: f_i^*(\mathbf{r}) f_j(\mathbf{r}) = \delta_{ij},
\end{equation}
the generalized Bloch susceptibility can be written as
\begin{equation}
    \bar{\chi}_{BA}(\mathbf{r},\mathbf{r}',\mathbf{q},z) = \sum_{i,j} f_i(\mathbf{r}) \chi_{BA}^{ij}(\mathbf{q},z) f_j^*(\mathbf{r}'),
\end{equation}
where
\begin{equation}
    \chi_{BA}^{ij}(\mathbf{q},z) = \iint_{\Omega_\mathrm{cell}}d\mathbf{r}d\mathbf{r}'\: f_i^*(\mathbf{r}) \bar{\chi}_{BA}(\mathbf{r},\mathbf{r}',\mathbf{q},z) f_j(\mathbf{r}').
\end{equation}
Noting that
\begin{align}\label{eq:note on matrix element conjugation}
    a_{\alpha'\alpha}^*(\mathbf{r}) &= \left[ \frac{\Omega}{\Omega_\mathrm{cell}} e^{i\mathbf{q}_{\alpha'\alpha}\cdot\mathbf{r}} \langle\alpha'|\hat{A}(\mathbf{r})|\alpha\rangle \right]^* 
    \nonumber \\
    &= \frac{\Omega}{\Omega_\mathrm{cell}} e^{-i\mathbf{q}_{\alpha'\alpha}\cdot\mathbf{r}} \langle\alpha|\hat{A}^\dagger(\mathbf{r})|\alpha'\rangle = a_{\alpha\alpha'}^\dagger(\mathbf{r}),
\end{align}
one can define basis projected matrix elements
\begin{equation}
    b_{\alpha\alpha'}^i \equiv \int_{\Omega_\mathrm{cell}} d\mathbf{r}\: f_i^*(\mathbf{r}) b_{\alpha\alpha'}(\mathbf{r})
\end{equation}
and use Eq. \eqref{eq:generalized bloch susceptibility lehmann} to write
\begin{align}
    \chi_{BA}^{ij}(\mathbf{q},z) = \frac{\Omega_\mathrm{cell}}{\Omega} \sum_{\alpha,\alpha'} 
    &b_{\alpha\alpha'}^i 
    \left[a^{\dagger\:j}_{\alpha\alpha'}\right]^* 
    \delta_{\mathbf{q},\mathbf{q}_{\alpha'\alpha}} 
    \nonumber \\
    &\times \frac{n_\alpha - n_{\alpha'}}{\hbar z - (E_{\alpha'} - E_\alpha)}.
\end{align}
In this representation, the reactive and dissipative parts can be expressed as
\begin{subequations}\label{eq:generalized bloch susceptibility reactive and dissipative parts}
    \begin{equation}
        \chi_{BA}'^{\:ij}(\mathbf{q},z) = \frac{1}{2}\left(\chi^{ij}_{BA}(\mathbf{q},z) + \left[\chi^{ji}_{A^\dagger B^\dagger}(\mathbf{q}, z)\right]^*\right)
    \end{equation}
    and
    \begin{equation}
        \chi_{BA}''^{\:ij}(\mathbf{q},z) = \frac{1}{2i}\left(\chi^{ij}_{BA}(\mathbf{q},z) - \left[\chi^{ji}_{A^\dagger B^\dagger}(\mathbf{q}, z)\right]^*\right),
    \end{equation}
\end{subequations}
where the basis-projected scattering function,
\begin{equation}
    S_{BA}^{ij}(\mathbf{q},z) = - \chi_{BA}''^{\:ij}(\mathbf{q},z) / \pi,
\end{equation}
in the limit of $\eta\rightarrow 0^+$ is given by
\begin{align}\label{eq:basis proj scattering function}
    S_{BA}^{ij}(\mathbf{q},\omega)
    = \frac{\Omega_\mathrm{cell}}{\Omega} \sum_{\alpha,\alpha'} 
    &b_{\alpha\alpha'}^i \left[a^{\dagger\:j}_{\alpha\alpha'}\right]^* \delta_{\mathbf{q},\mathbf{q}_{\alpha'\alpha}}
    \nonumber \\
    &\times (n_\alpha - n_{\alpha'})\; \delta\hspace{-1pt}\left(\hbar \omega - [E_{\alpha'} - E_\alpha]\right).
\end{align}
In this sum over states, each unique pair of eigenstates $|\alpha\rangle$ and $|\alpha '\rangle$ contribute twice. Assuming that each basis function's complex conjugate itself is included in the basis (e.g. plane waves or a real set of basis functions), one may denote $f_{i^*}(\mathbf{r})=f_i^*(\mathbf{r})$ and use Eq. \eqref{eq:note on matrix element conjugation} to write
\begin{equation}
    b_{\alpha'\alpha}^i = \int_{\Omega_\mathrm{cell}} d\mathbf{r}\: f_{i^*}(\mathbf{r}) \left[b^\dagger_{\alpha\alpha'}(\mathbf{r})\right]^* = \left[b_{\alpha\alpha'}^{\dagger i^*}\right]^*.
\end{equation}
Insertion into Eq. \eqref{eq:basis proj scattering function} then allows for seamless separation of the two contributions in terms of spectral functions
\begin{equation}\label{eq:basis proj scattering function separation}
    S_{BA}^{ij}(\mathbf{q},\omega) = A_{BA}^{ij}(\mathbf{q},\omega) - A_{AB}^{j^*i^*}(-\mathbf{q},-\omega),
\end{equation}
where
\begin{align}\label{eq:basis proj spectral function}
    A_{BA}^{ij}(\mathbf{q},\omega) \equiv \frac{\Omega_\mathrm{cell}}{\Omega} \sum_{\alpha'>\alpha} 
    &b_{\alpha\alpha'}^i \left[a^{\dagger\:j}_{\alpha\alpha'}\right]^* \delta_{\mathbf{q},\mathbf{q}_{\alpha'\alpha}}
    \nonumber \\
    &\times (n_\alpha - n_{\alpha'})\; \delta\hspace{-1pt}\left(\hbar \omega - [E_{\alpha'} - E_\alpha]\right),
\end{align}
and the eigenstate index $\alpha$ is assumed ordered by energy (ascending) and occupation (descending), such that each spectral function $A_{BA}^{ij}(\mathbf{q},\omega)$ has spectral weight only at nonnegative frequencies.  
Finally, one may from Eq. \eqref{eq:generalized bloch susceptibility reactive and dissipative parts} note that the reactive and dissipative parts of $\chi_{A^\dagger A}^{ij}(\mathbf{q},z)$ simply become the Hermitian and anti-Hermitian parts of the $(i,j)$ matrix, while the corresponding spectral function, $A_{A^\dagger A}^{ij}(\mathbf{q},z)$, is Hermitian and positive-semidefinite.

\section{Convergence of the magnon dispersion}\label{app:magnon dispersion convergence}

\subsection{Basis convergence}\label{sec:pw convergence}

\begin{figure*}[tb]
    \centering
    \includegraphics[scale=1.0]{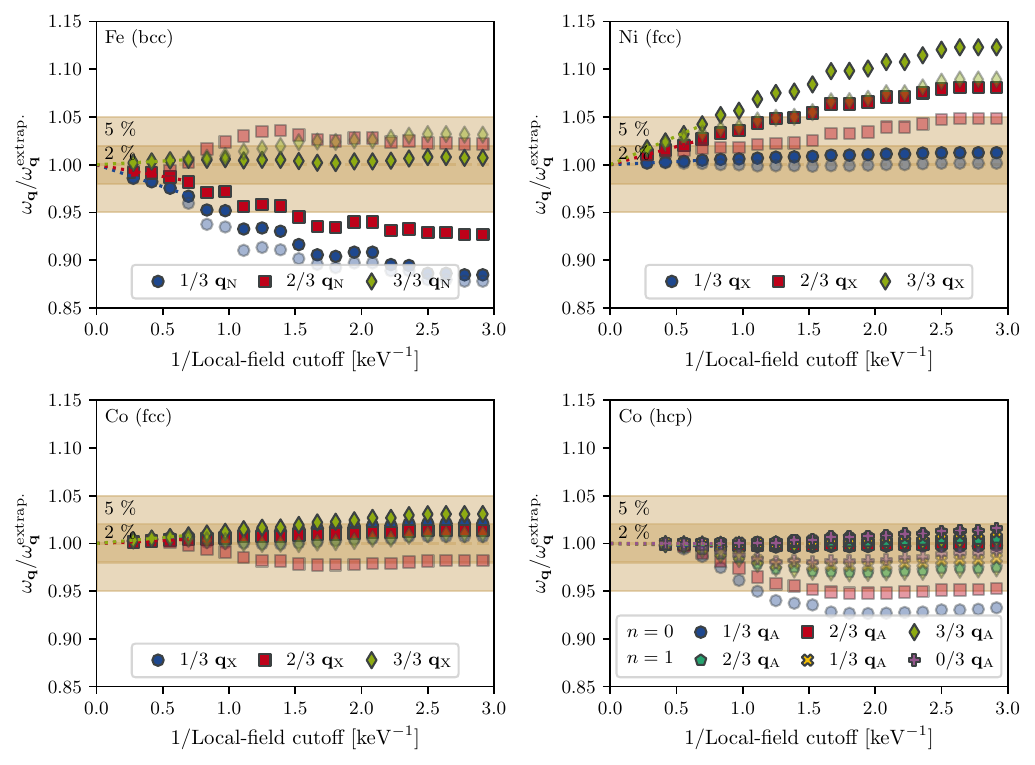}
    \caption{Magnon peak frequency convergence as a function of the inverse local-field cutoff, including 32 empty-shell bands per atom. For hcp-Co, $n=0$ indicate the acoustic (Goldstone) magnon mode, while $n=1$ indicate the optical mode. The transparency of the markers indicates the gap error compensation scheme applied; $\Xi^{++}$-rescaled $\omega_\mathbf{q}$ are opaque while rigidly shifted $\omega_\mathbf{q}$ are translucent. To illustrate the relative error at finite cutoffs, the magnon frequencies are plotted relative to a linear extrapolation of the $\Xi^{++}$-rescaled dispersion to infinite plane-wave cutoffs (dotted lines), see also Tab. \ref{tab:extrapolated magnon frequencies}.}
    \label{fig:magnon_dispersion_ecut_convergence}
\end{figure*}
%
In order to systematically investigate the convergence of the magnon dispersion in the limit of a complete basis representation, 
we use 32 empty-shell bands per atom, $\Gamma$-centered $k$-point grids of size $90\times90\times90$ for bcc-Fe, $84\times84\times84$ for fcc-Ni and fcc-Co, and $60\times60\times36$ for hcp-Co (the same $k$-point grids were used in the gap error calculations above), along with a finite broadening of $\eta=50$ meV in Eqs. \eqref{eq:plane-wave Kohn-Sham susceptibility} and \eqref{eq:plane-wave self-enhancement function} to robustly resolve the magnon peak positions \cite{Skovhus2021}. 
In Fig. \ref{fig:magnon_dispersion_ecut_convergence} we present the resulting plane-wave convergence of the magnon dispersion in Fe, Ni and Co, extracting the magnon peak frequencies $\omega_\mathbf{q}$ from the 
majority magnon lineshapes \eqref{eq:majority magnon lineshape}.
%
For plane-wave cutoffs lower than 2 keV, the gap error is non-negligible and needs to be compensated by a correction. In Fig. \ref{fig:magnon_dispersion_ecut_convergence} we investigate two such gap error compensation schemes; rescaling $\Xi^{++}\rightarrow\lambda \Xi^{++}$ to satisfy the Goldstone criterion \eqref{eq:Goldstone criterion} and rigidly shifting the entire magnon spectrum $\omega_\mathbf{q}\rightarrow\omega_\mathbf{q}-\omega_\Gamma$. For fcc-Co, both strategies fully compensate for the gap error such that the magnon dispersion becomes converged within 5\% even for plane-wave cutoffs below 400 eV. For hcp-Co, the $\Xi^{++}$-rescaling performs even better still with relative errors below 2\%, in contrast to the rigid shift, which is less effective than for fcc-Co. For fcc-Ni, the trend is opposite. Here, the errors are smallest when applying a rigid shift to the magnon frequencies, whereas the relative performance of the two compensation schemes depend on the selected wave vector $\mathbf{q}$ in bcc-Fe. 
%

Even though the results in Fig. \ref{fig:magnon_dispersion_ecut_convergence} are inconclusive concerning which compensation scheme yields the lowest overall error, we observe that the magnon peak frequencies generally depend more monotonously on the inverse local-field cutoff when $\Xi^{++}$ is rescaled to meet the Goldstone criterion \eqref{eq:Goldstone criterion}. This is likely due to an improved alignment of the magnon resonances to the Stoner continuum governed by the Kohn-Sham excitation energies in the denominator of Eqs. \eqref{eq:plane-wave Kohn-Sham susceptibility} and \eqref{eq:plane-wave self-enhancement function}. Whereas the spectral features of the Stoner continuum are unaffected by the plane-wave cutoff, it is the variation in collective resonance frequencies [the frequencies at which $\mathrm{Re}\:\langle v_n(\mathbf{q})|\Xi^{++}(\mathbf{q},\omega)|v_n(\mathbf{q})\rangle=1$] that drives the plane-wave convergence in Figs. \ref{fig:gap_error_convergence} and \ref{fig:magnon_dispersion_ecut_convergence}. In the ideal case, the error on the collective resonances (due to the truncated plane-wave representation) is $\mathbf{q}$-independent. However, even in this case, a rigid shift of $\omega_\mathbf{q}\rightarrow\omega_\mathbf{q}-\omega_\Gamma$ cannot correct for compounding errors that are picked up through the hybridization of the misaligned magnon resonance with the correctly aligned Stoner continuum. 
As the gap error vanishes with increasing plane-wave cutoff, the magnon resonances at finite $\mathbf{q}$ may pass through various spectral features in the Stoner spectrum and through the hybridization with these pick up a nontrivial (nonmonotonous) plane-wave dependence. By rescaling $\Xi^{++}\rightarrow\lambda \Xi^{++}$, one hopes to correct not only the gap error at the $\Gamma$-point, but also the spectral alignment of collective resonances at finite $\mathbf{q}$, thereby providing a more monotonous plane-wave convergence.
%
%
\begin{table}[tb]
    \centering
    \begin{tabular}{c c|c|c|c|c} 
         & &
        Fe (bcc) & Ni (fcc) & Co (fcc) & Co (hcp) \\ 
        \hline
        \multirow{3}{.25cm}{\rotatebox[origin=c]{90}{$n=0$}} & $1/3$ $\mathbf{q}_s$ &
        $52.5 \pm 0.1$ & $308.8 \pm 0.1$ & $160.1 \pm 0.1$ & $36.55 \pm 0.01$ \\
        & $2/3$ $\mathbf{q}_s$ &
        $178.4 \pm 0.5$ & $497.3 \pm 0.6$ & $486.4 \pm 0.1$ & $132.93 \pm 0.02$ \\
        & $3/3$ $\mathbf{q}_s$ &
        $336.4 \pm 0.1$ & $477.2 \pm 0.9$ & $763.4 \pm 0.1$ & $296.61 \pm 0.03$ \\
        \hline
        \multirow{3}{.25cm}{\rotatebox[origin=c]{90}{$n=1$}} & $2/3$ $\mathbf{q}_s$ & 
        & & & $438.2 \pm 0.1$ \\
        & $1/3$ $\mathbf{q}_s$ & 
        & & & $532.8 \pm 0.2$ \\
        & $0/3$ $\mathbf{q}_s$ & 
        & & & $558.2 \pm 0.3$ \\
    \end{tabular}
    \caption{Extrapolated ALDA magnon frequencies $\hbar\omega_\mathbf{q}^\mathrm{extrap.}$ in meV for wave vectors $\mathbf{q}$ along the high-symmetry path from the $\Gamma$-point to high-symmetry points $s=\mathrm{N}$ (bcc), $s=\mathrm{X}$ (fcc) and $s=\mathrm{A}$ (hcp), see also Fig. \ref{fig:magnon_dispersion_ecut_convergence}. The tabulated uncertainty is the standard deviation of $\hbar\omega_\mathbf{q}^\mathrm{extrap.}$ in the linear fit used for the extrapolation.}
    \label{tab:extrapolated magnon frequencies}
\end{table}
Inspired by the observed monotony, we fit the $\Xi^{++}$-rescaled magnon frequencies in Fig. \ref{fig:magnon_dispersion_ecut_convergence} to a linear function of the inverse local-field cutoff (using the four highest cutoffs available for each crystal). With this fit in hand, we extrapolate the magnon frequencies to an infinite plane-wave cutoff. The extrapolated frequencies are listed in Tab. \ref{tab:extrapolated magnon frequencies} and the plane-wave convergence in Fig. \ref{fig:magnon_dispersion_ecut_convergence} is plotted relative to these values. By the virtue of basis completeness, we believe that this data set provides a valuable benchmark for future implementations, especially implementations with less general periodic basis representations. For the plane-wave implementation considered here, we note that it generally is more difficult to converge the magnon dispersion of Fe and Ni compared to Co, see Fig. \ref{fig:magnon_dispersion_ecut_convergence}. As a result, we use a plane-wave cutoff of 1.5 keV for Fe and Ni in the calculations of Sec. \ref{sec:transition metals results}, while 1.0 keV of plane-waves are included in the basis representation of the Co crystals.

\subsection{Band convergence}\label{sec:band convergence}

\begin{figure*}[tb]
    \centering
    \includegraphics[scale=1.0]{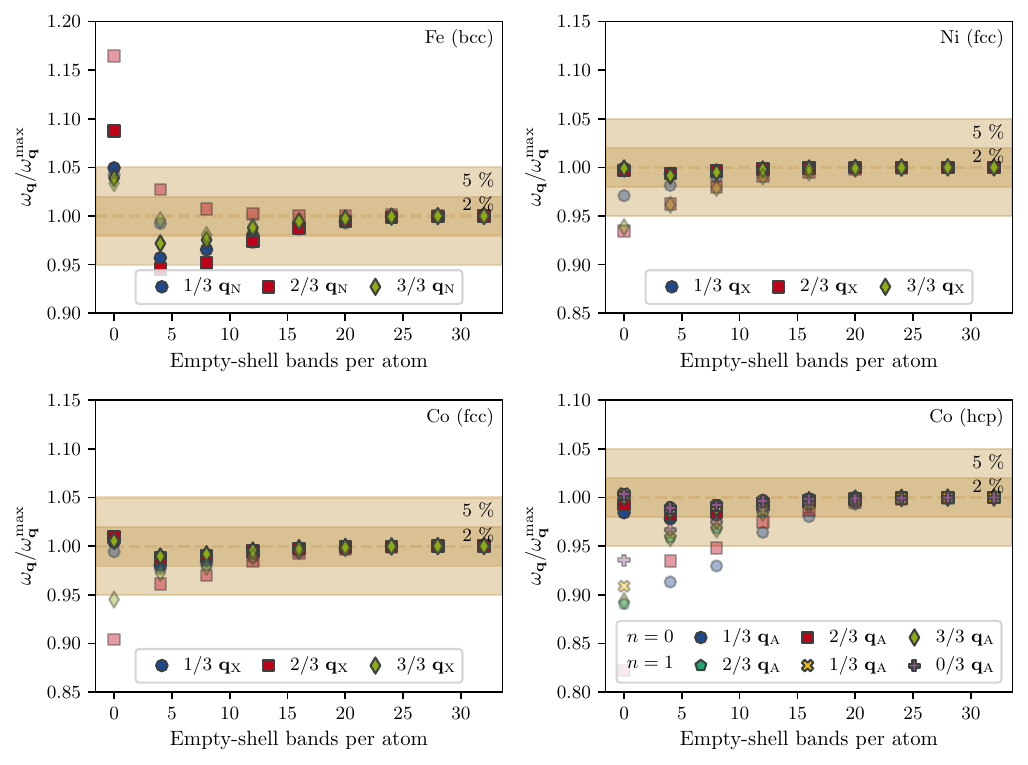}
    \caption{Magnon peak frequency convergence as a function of the number of empty-shell bands per atom, inverting the Dyson Eq. \eqref{eq:ALDA dyson equation} using a plane-wave cutoff of 1.5 keV for Fe/Ni and 1.0 keV for Co. For hcp-Co, $n=0$ indicate the acoustic (Goldstone) magnon mode, while $n=1$ indicate the optical mode. The transparency of the markers indicates the gap error compensation scheme applied; $\Xi^{++}$-rescaled $\omega_\mathbf{q}$ are opaque while rigidly shifted $\omega_\mathbf{q}$ are translucent. To illustrate the relative error at finite cutoffs, the magnon frequencies are plotted relative to the frequency at 32 empty-shell bands per atom.}
    \label{fig:magnon_dispersion_unocc_convergence}
\end{figure*}
Even though it is quite feasible to converge away the gap error arising from the finite number of bands included in Eqs. \eqref{eq:plane-wave Kohn-Sham susceptibility} and \eqref{eq:plane-wave self-enhancement function}, see Fig. \ref{fig:gap_error_convergence}, there is still a lot of computational time to be saved from an effective gap error compensation scheme (especially for more complex crystals than the ones considered here). In Fig. \ref{fig:magnon_dispersion_unocc_convergence}, we investigate the band convergence of the magnon dispersion in Fe, Ni and Co when applying $\Xi^{++}$-rescaling versus a rigid shift to compensate for the gap error. Although both compensation schemes improve convergence, it seems that rescaling $\Xi^{++}$ to meet the Goldstone criterion \eqref{eq:Goldstone criterion} is the most effective strategy to compensate for the finite number of bands. When applying a rescaling to the self-enhancement function of Ni and Co, the magnon peak frequencies can be converged within 2\% without any empty-shell bands, despite bare gap errors on the order of 50-100 meV. For Fe, band convergence is more tricky and 16 empty-shell bands per atom are needed to converge the magnon dispersion within 2\% with either gap error compensation scheme. The reason for this is unclear, but likely related to the coupling between magnon resonances and Stoner continuum discussed above. For the calculations of the full magnon spectrum in Fe, Ni and Co presented in Sec. \ref{sec:transition metals results}, we use 12 empty-shell bands per atom for Ni and Co while including an additional 4 for Fe. 
In total, the bare gap errors with the chosen number of bands and plane-wave cutoffs are 12.9 meV for Fe (bcc), 12.8 meV for Ni (fcc), 32.8 meV for Co (fcc) and 31.4 meV for Co (hcp). In order to satisfy the Goldstone criterion \eqref{eq:Goldstone criterion} for the presented magnon spectra, the self-enhancement function is rescaled $\Xi^{++}\rightarrow\lambda \Xi^{++}$ with $\lambda=1.0054$ for Fe (bcc), $\lambda=1.0162$ for Ni (fcc), $\lambda=1.0177$ for Co (fcc) and $\lambda=1.0171$ for Co (hcp). Finally, we use $\Gamma$-centered $k$-point grids sized $88 \times 88 \times 88$ for the bcc and fcc structures and $66 \times 66 \times 36$ for hcp-Co along with an artificial broadening of 50 meV to reliably resolve the spectral features of $\chi^{+-}(\mathbf{q},\omega)$ in Sec. \ref{sec:transition metals results} for a wide range of commensurate $\mathbf{q}$-points.

\section{Mode decomposition in Fe and Ni}\label{app:mode decomposition}

\begin{figure*}[tb]
    \centering
    \includegraphics[scale=1.0]{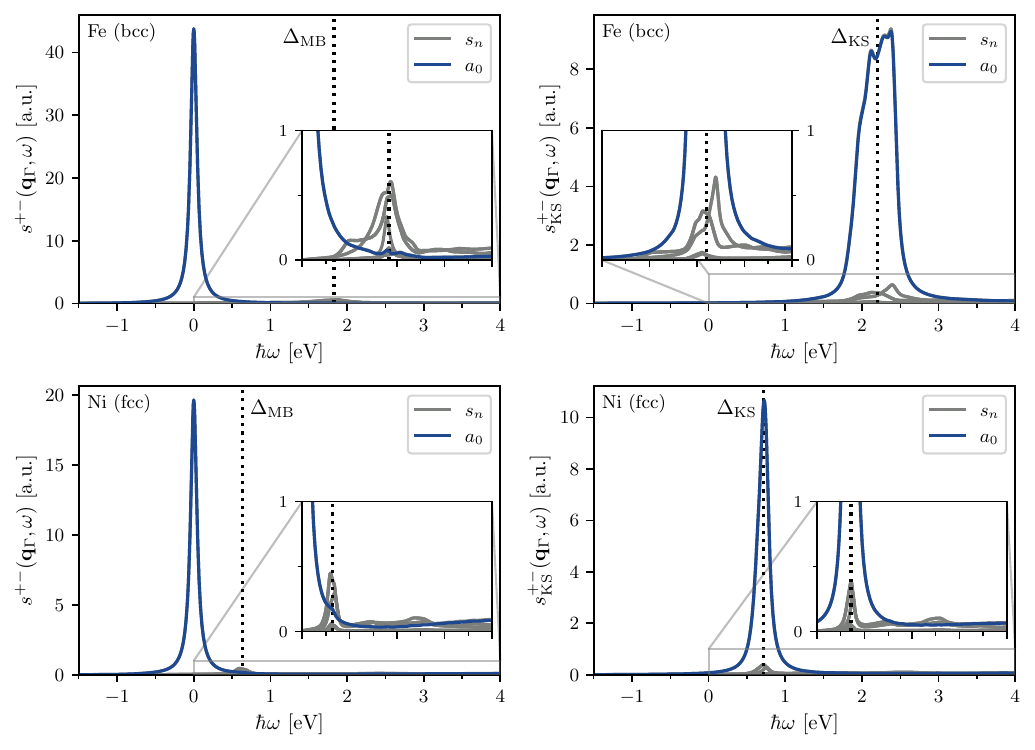}
    \caption{Eigenvalues $s^{+-}_n$ (grey) of the scattering function \eqref{eq:bloch scattering function eigendecomposition} in Fe (top) and Ni (bottom) at zero momentum transfer for the many-body system (left) and Kohn-Sham system (right). The 12 largest eigenvalues are plotted. In blue, the projection \eqref{eq:majority magnon lineshape} of the majority spectral function \eqref{eq:bloch spectral function reconstructed} onto the extracted Goldstone mode $|v_0(\mathbf{q}_\Gamma)\rangle$ is shown, lying virtually on top of the corresponding eigenvalue for a wide range of frequencies. The effective (Kohn-Sham) exchange splitting $\Delta_\mathrm{KS}$ and (many-body) Stoner pair gap $\Delta_\mathrm{MB}$---see Tab. \ref{tab:stoner shift}---are shown as vertical dotted lines.}
    \label{fig:FeNi_eigendecomposition}
\end{figure*}
In Fig. \ref{fig:FeNi_eigendecomposition} the eigenvalues of the scattering function \eqref{eq:bloch scattering function} are show for Fe (bcc) and Ni (fcc), complementing Fig. \ref{fig:Co_eigendecomposition} for Co (fcc/hcp). The discussion of collective versus single-particle modes of Sec. \ref{sec:mode decomposition} clearly applies here as well. The magnon resonance is entirely contained within a single eigenmode of the many-body scattering function $S^{+-}$, whereas the rest of the eigenmodes make out the many-body continuum of Stoner pair excitations. Furthermore, the collective mode perfectly overlaps with an eigenmode of the Kohn-Sham scattering function $S_\mathrm{KS}^{+-}$, whose entire spectral weight is redistributed to the magnon resonance by the Dyson Eq. \eqref{eq:ALDA dyson equation}.

In addition to the bare eigenvalues, we show in Figs. \ref{fig:Co_eigendecomposition} and \ref{fig:FeNi_eigendecomposition} also the magnon lineshapes computed by projecting the majority spectral function onto the extracted magnon mode vectors. In this regard, a few comments are in order. Firstly, we do not compute all frequencies for a given $\mathbf{q}$-point in one go. This means that we need to store the scattering function $S^{+-}(\mathbf{q},\omega)$ on disk and extract the magnon mode vectors $|v_n(\mathbf{q})\rangle$ later (in order to optimize the extraction frequency $\omega$). However, it is not feasible to store the entire scattering function for all crystal momenta and frequencies of interest, so after inverting the Dyson Eq. \eqref{eq:ALDA dyson equation}, we reduce the plane-wave basis of $\chi^{+-}(\mathbf{q},\omega)$ to a plane-wave cutoff of 250 eV.
In this reduced basis, we compute the scattering function \eqref{eq:bloch scattering function} and construct the majority spectral function $A^{+-}(\mathbf{q},\omega)$ from its positive eigenvalues \eqref{eq:bloch spectral function reconstructed}.
%
%
With the majority spectral function in hand, we save its full trace along with the 12 or 24 (bcc/fcc or hcp) largest eigenvalues and corresponding eigenvectors. This way, we can reconstuct it approximately at any later time (e.g. when all calculations are finished). Except for Fig. \ref{fig:total_scattering}---which shows the full trace of $S^{+-}(\mathbf{q},\omega)$ in the original 1000/1500 eV basis---all other results in Sec. \ref{sec:transition metals results} were obtained using the truncated eigendecomposition of $A^{+-}(\mathbf{q},\omega)$ and $A_\mathrm{KS}^{+-}(\mathbf{q},\omega)$ in the reduced plane-wave basis. Similarly, also the shown projections of the self-enhancement function $\Xi^{++}(\mathbf{q},\omega)$ are calculated using a truncated eigendecomposition in the 250 eV basis, in this case using the 22 or 44 (bcc/fcc or hcp) largest eigenvalues and corresponding eigenvectors.

\bibliography{bibliography}

\end{document}